# Toward a muon-specific electronic structure theory: Effective electronic Hartree-Fock equations for the muonic molecules

Milad Rayka[1], Mohammad Goli[2,*] and Shant Shahbazian[1,*]

[1] *Department of Physics and Department of Physical and Computational Chemistry, Shahid Beheshti University, G. C., Evin, Tehran, Iran, 19839, P.O. Box 19395-4716.*

[2] *School of Nano Science, Institute for Research in Fundamental Sciences (IPM), Tehran 19395-5531, Iran*

E-mails:

Mohammad Goli: mgoli2019@gmail.com

Shant Shahbazian: sh_shahbazian@sbu.ac.ir

[*] Corresponding authors

## Abstract

An effective set of the Hartree-Fock (HF) equations are derived for electrons of the muonic systems, i.e., molecules containing a positively charged muon, conceiving the muon as a quantum oscillator, which are completely equivalent to the usual two-component HF equations used to derive stationary states of the muonic molecules. In these effective equations, a non-Coulombic potential is added to the orthodox coulomb and exchange potential energy terms, which describes the interaction of the muon and the electrons effectively and is optimized during the self-consistent field cycles. While in the two-component HF equations muon is treated as a quantum wave, in the effective HF equations it is absorbed into the effective potential and practically transformed into an effective potential field experienced by electrons. The explicit form of the effective potential depends on the nature of muon's vibrations and is derivable from the basis set used to expand the muonic spatial orbital. The resulting effective Hartree-Fock equations are implemented computationally and used successfully, as a proof of concept, in a series of the muonic molecules containing all atoms from the second and third rows of the Periodic Table. To solve the algebraic version of the equations muon-specific gaussian basis sets are designed for both muon and surrounding electrons and it is demonstrated that the optimized exponents are quite distinct from those derived for the hydrogen isotopes. The developed effective HF theory is quite general and in principle can be used for any muonic system while it is the starting point for a general effective electronic structure theory that incorporates various types of quantum correlations into the muonic systems beyond the HF equations.

## Keywords

Isotope effects; ab initio electronic structure calculations; Positively charged muon; non-adiabatic effective Hamiltonian

## I. Introduction

The conventional electronic structure theory of molecules is deeply rooted in the adiabatic paradigm; electrons are considered as quantum particles governed by the electronic Schrödinger equation while nuclei are treated at first step as clamped point charges.[1-3] In next step the electronic energies derived from the electronic Schrödinger equation are used to construct the potential energy surfaces, which are employed as effective potentials for the nuclear Schrödinger equation.[4-7] This two-step procedure and the separation of Schrödinger equation into electronic and nuclear equations is in the heart of modern quantum chemistry while non-adiabatic effects, which are beyond this scheme, when needed to be taken into account, are generally treated by special techniques as small perturbations.[8] In contrast to the significant success of this paradigm, the separate consideration of electrons and nuclei is just an approximation, albeit a good one, but not an inherent trait of molecular quantum mechanics. Accordingly, a fully non-adiabatic paradigm has been developed and implemented computationally in recent decades that bypasses the conventional adiabatic paradigm and treats electrons and nuclei equally from the outset as quantum particles within the context of a single Schrödinger equation.[9-12] In this paradigm the molecular Schrödinger equation is solved using the variational based techniques and for few-particle systems,[13] highly accurate numerical solutions are derived, which may be conceived as exact for all practical purposes. However, the used computational methods are quite computationally demanding and in their original formulation,[9-13] cannot probably be extended in foreseeable future beyond few-particle systems. To have less computationally demanding methods, there have been independent attempts to extend the molecular orbital approach within the non-adiabatic paradigm while the terminology and certain technical details are varied, the general idea is attributing orbitals to both electrons and nuclei.[14-18] The resulting orbital-based ab initio procedures are similar to their adiabatic counterparts, Hartree-Fock (HF), post-HF methods and density functional theory (DFT), all concentrating on accurate computation of the electron-electron correlation but usually neglect or treat the electron-nucleus correlation improperly. However, more recently, it has been demonstrated that accurate evaluation of the electron-nucleus correlation is crucial for quantitative reproduction of the nuclear vibrational frequencies.[19] To evaluate this type of correlation there has been some attempts to design electron-nucleus functionals within the context of the non-adiabatic DFT,[20-29] as well as trying to use the explicitly correlated electron-nucleus Gaussian geminals.[19,30-37] Though both approaches are leaps forward, a theoretically justified and at the same time computationally tractable method to accurately evaluate the electron-nucleus correlation seems to be elusive yet and much remains to be done in this area. It is timely to emphasize that in practice one rarely needs to consider all nuclei as quantum particles and a mixed intermediate adiabatic/non-adiabatic paradigm is a proper compromise; certain light nuclei, e.g. protons and its heavier isotopes, are considered as quantum particles while all the heavier nuclei are conceived as clamped point charges. Although there are certain non-adiabatic process involving protons, which must be considered inherently within context of the non-adiabatic paradigm, e.g. proton-coupled electron transfer,[38] most applications of orbital-based non-adiabatic ab initio procedures were centered around various hydrogen isotope effects, which are not inherently non-adiabatic. It is worth noting that even without including electron-nucleus correlation at all, various hydrogen isotope effects have been considered using the orbital-based non-adiabatic ab initio procedures and most (but not all) results seem to agree with the known observed experimental trends (See Ref. 53 for a coverage of the relevant literature before 2011).[39-52] This success is by part understandable based on the fact that the electron-nucleus correlation is seemingly local by its nature and relatively (in comparison to the electron-electron correlation) insensitive to its environment. Thus, while the mentioned ab initio procedures are probably unable to reproduce absolute values of the nuclear vibrational frequencies, they may reproduce the correct trends in isotopically substituted series of molecules. The use of orbital-based non-adiabatic ab initio procedures in cases and processes that are not inherently non-adiabatic is justified computationally since the alternative procedure within the adiabatic paradigm namely, constructing a potential energy surface to consider the quantum nature of nuclei is usually computationally more demanding. Accordingly, the current usage of the orbital-based non-adiabatic ab initio procedures to consider the isotopic and other subtle effects,[54,55] must be seen as an alternative that tries to mimic the solutions of the nuclear vibrational Schrödinger equation as accurately as possible. It still remains to be seen whether one may design an orbital-based non-adiabatic ab initio procedure which be capable of accurately reproducing both adiabatic and the purely non-adiabatic isotope effects simultaneously.

While employing the orbital-based non-adiabatic ab initio procedures are not obligatory for most processes involving protons and its heavier isotopes, in the case of muonic molecules, i.e. molecules containing the positively charged muons, which is only ~206 time heavier than electron, the adiabatic paradigm is not always a safe approximation and the use of non-adiabatic ab initio procedures are much more justified. Indeed, recent progress in the muon spin resonance spectroscopy opens a new window into the reaction of the muonium atom,[56-63] i.e. an exotic atom composed of a positively charged muon and an electron, with various organic molecules and the resulting organic radicals are of prime interest.[64-75] In order to consider the geometry, the sticking site of the muon and the electronic structure of the muonic molecules various theoretical and computational methods have been proposed in the last three decades.[76-85] Recently, we have also employed the Nuclear-Electronic Orbital (NEO) non-adiabatic ab initio procedure,[14,86] for the muonic molecules and the primary results were promising though our treatment lacked the electron-muon correlation.[87-90] Since the NEO theory has been originally formulated within the mixed intermediate adiabatic/non-adiabatic paradigm,[14,86] i.e. only selected nuclei are treated as a quantum particles instead of clamped point charges, the NEO is an ideal framework for ab initio calculations on the muonic molecules. The Schrödinger equation within context of the NEO contains the kinetic energy operators of electrons and the muon, and at the NEO-HF computational level the wavefunction is approximated as a product of a Slater determinant for electrons and a spin-orbital for the muon; the resulting coupled HF equations are solved simultaneously for electrons and the muon.[87-89] One may try to incorporate the electron-electron and electron-muon correlations into the NEO methodology, which are both absent at the NEO-HF level. This can be done systematically through employing more complicated wavefunctions than that of the NEO-HF, using the variational principle or the perturbation theory;[14,91] introducing electron-muon correlation is a subtle step and in the meantime, it seems more legitimate to add the electron-muon correlation through using simplified physical models as will be considered in a future study. The second-order Møller-Plesset perturbation correction, with the NEO-HF wavefunction as the reference unperturbed wavefunction, has been recently implemented computationally for the muonic systems.[90] In principle, it is possible to employ the full arsenal of the NEO methodology for the muonic molecules, however, based on a recent proposed alternative effective formulation of the NEO-HF method, called the effective HF theory (EHF),[92] it seems feasible to introduce a *muon-specific electronic structure theory*. The main ingredient behind the EHF theory is the use of the Hartree product nuclear wavefunction of nuclear spin-orbitals instead of the Slater determinant, which was proposed originally by Nakai,[93] and then employed by Auer and Hammes-Schiffer in their formulation of the nuclear exchange-free NEO-HF equations.[94] This replacement was justified since nuclei are intrinsically localized and the overlap of the nuclear orbitals, in most situations, are negligible promising that the nuclear exchange integrals are practically null and thus the nuclei are "distinguishable" particles. *The EHF theory eliminates the nuclear spatial orbitals from the variational integral by direct integration on the nuclear spatial variables and transforming the nuclear orbitals to "parameters" in an effective non-Coulombic potential experienced by electrons*. Eventually, the *optimized* effective potential and the electronic orbitals are both determined through the self-consistent field (SCF) procedure, which enables one to reproduce the nuclear orbitals. In this paper, the effective EHF equations are explicitly introduced for the muonic molecules as an alternative to the NEO-HF equations. This is the first ladder in the construction of the muon-specific electronic structure theory and its computational implementation for ab initio calculations. In order to perform ab initio calculations, muonic basis sets are required and according to the best of our knowledge no systematic study has been done yet to design *muon-specific basis sets*. Thus, novel muonic and corresponding electronic gaussian basis sets are also designed for both muon and surrounding electrons, respectively, employing protocols developed by Hammes-Schiffer and coworkers,[14,95] and by Tachikawa and coworkers,[96-102] which were used previously to design energy-optimized basis sets for the hydrogen isotopes. The resulting basis sets are then used for ab initio EHF calculations on a series of muonic species in conjunction with conventional electronic basis sets, i.e. designed within the adiabatic paradigm, revealing the need for carful design of the muon-specific basis sets.

## II. The EHF theory

The details of deriving the EHF equations as well as its conceptual implications have been discussed previously thus only the main steps are considered herein.[92] The NEO Hamiltonian for a muonic system, containing $N_e$ electrons, a single muon and $q$ clamped nuclei, written in atomic units is as follows:

$$\hat{H}^{\mu}_{total} = \hat{H}^{\mu}_{NEO} + \sum_{\beta}^{q}\sum_{\gamma\rangle\beta}^{q}\frac{Z_\beta Z_\gamma}{\left|\vec{R}_\beta - \vec{R}_\gamma\right|}$$

$$\hat{H}^{\mu}_{NEO} = \left(-1/2\right)\sum_{i}^{N_e}\nabla_i^2 + \left(-1/2\,m_\mu\right)\nabla_\mu^2 - \sum_{i}^{N_e}\frac{1}{\left|\vec{r}_\mu - \vec{r}_i\right|}$$

$$+\sum_{i}^{N_e}\sum_{j\rangle i}^{N_e}\frac{1}{\left|\vec{r}_i - \vec{r}_j\right|} - \sum_{i}^{N_e}\sum_{\beta}^{q}\frac{Z_\beta}{\left|\vec{R}_\beta - \vec{r}_i\right|} + \sum_{\beta}^{q}\frac{Z_\beta}{\left|\vec{R}_\beta - \vec{r}_\mu\right|} \quad (1)$$

In order to employ the variational principle, the following trial normalized wavefunction: $\Psi_{trial} = \psi_e(\vec{r}_1,...,\vec{r}_{N_e})\psi_\mu\left(\vec{r}_\mu\right)$ (the spin variables have been neglected for brevity) is proposed, which at this stage contains the full electron-electron correlation in principle but neglects the electron-muon correlation from the outset (for a similar idea albeit in a different context see Ref. 103 and 104). Incorporating the trial ground state wavefunction into the variational integral one arrives at:

$$\mathrm{E} = \int ... \int d\vec{r}_1..d\vec{r}_{N_1}\ \psi_e^*(\vec{r}_1,...,\vec{r}_{N_e})\ \hat{H}_e^{eff}\ \psi_e(\vec{r}_1,...,\vec{r}_{N_e})$$

$$\hat{H}_e^{eff} = \left(-1/2\right)\sum_{i}^{N_e}\nabla_i^2 + \sum_{i}^{N_e}\sum_{j\rangle i}^{N_e}\frac{1}{\left|\vec{r}_i - \vec{r}_j\right|} - \sum_{i}^{N_1}\sum_{\beta}^{q}\frac{Z_\beta}{\left|\vec{R}_\beta - \vec{r}_i\right|} + V^{eff}$$

$$V^{eff} = \left(-1/2m_\mu\right)\int d\vec{r}_\mu \psi_\mu^* \nabla_\mu^2 \psi_\mu - \sum_{i}^{N_e}\int d\vec{r}_\mu \frac{\psi_\mu^*\psi_\mu}{\left|\vec{r}_\mu - \vec{r}_i\right|} + \sum_{\beta}^{q} Z_\beta \int d\vec{r}_\mu \frac{\psi_\mu^*\psi_\mu}{\left|\vec{R}_\beta - \vec{r}_\mu\right|} \quad (2)$$

At this stage of development, one may claim that the muon has been effectively transformed from a quantum particle with dynamical variables into an effective potential, which is experienced by electrons. The simplest possible representation of the muonic orbital is based on a single s-type gaussian function, $\psi_{\mu-s} = (2\alpha/\pi)^{\frac{3}{4}} Exp\left(-\alpha\left|\vec{r}_\mu - \vec{R}_c\right|^2\right)$, where $\vec{R}_c = \hat{i}X_c + \hat{j}Y_c + \hat{k}Z_c$, which has been used previously in ab initio NEO-HF calculations as a muonic basis set.[87-90] The s-type gaussian function is used herein to convey the main ingredients of the EHF equations while more flexible basis set will be considered in subsequent section (the subscript $s$ is used throughout the paper to stress that the corresponding functions are derived from the s-type gaussian).[87-90] Incorporating this function into the general expression of the effective potential given in equation (2), and after some mathematical manipulations, the following effective potential is derived:

$$V_s^{eff} = V_{e-s}^{eff} + U_s^{eff}$$

$$V_{e-s}^{eff} = -\sum_i^{N_e} \frac{1}{\left|\vec{r}_i - \vec{R}_c\right|} erf\left[\sqrt{2\alpha}\left|\vec{r}_i - \vec{R}_c\right|\right]$$

$$U_s^{eff} = \left(\frac{3\alpha}{2m_\mu}\right) + \sum_\beta^q \frac{Z_\beta}{\left|\vec{R}_\beta - \vec{R}_c\right|} erf\left[\sqrt{2\alpha}\left|\vec{R}_\beta - \vec{R}_c\right|\right] \tag{3}$$

It is evident that $V_{e-s}^{eff}$ is non-Coulombic; the error function damps the Coulombic terms near the center of the gaussian function, $\vec{R}_c$. Through incorporating the derived $V_s^{eff}$ into the variational integral, equation (2), and employing the variational principle, $\delta_{\psi_e,\{\alpha,\vec{R}_c\}} \mathrm{E} = 0$, both electronic wavefunction and the energy-optimized parameters of the muonic orbital, i.e. $\alpha$ and $\vec{R}_c$, are derivable. The simplest case is to neglect the electron-electron correlation and employ a Slater determinant for the trial electronic wavefunction then performing the usual functional variational as is done to derive the conventional HF equations.[1] The resulting EHF equations for a closed electronic shell are as follows:

$$\hat{f}_s^{eff}\left(\vec{r}_1\right)\psi_i\left(\vec{r}_1\right)=\varepsilon_i\psi_i\left(\vec{r}_1\right) \qquad i=1,\dots,N_e/2$$

$$\hat{f}_s^{eff}\left(\vec{r}_1\right)=\hat{h}\left(\vec{r}_1\right)-\frac{1}{\left|\vec{r}_1-\vec{R}_c\right|}erf\left[\sqrt{2\alpha}\left|\vec{r}_1-\vec{R}_c\right|\right]+\sum_j^{N_e/2}\left[2\hat{J}_j\left(\vec{r}_1\right)-\hat{K}_j\left(\vec{r}_1\right)\right]$$

$$\hat{h}\left(\vec{r}_1\right)=\left(-1/2\right)\nabla_1^2-\sum_\beta^q\frac{Z_\beta}{\left|\vec{r}_1-\vec{R}_\beta\right|} \qquad (4)$$

In these equations, $\hat{J}_j$ and $\hat{K}_j$ are the usual coulomb and exchange operators, respectively,[1] while $\psi_i\left(\vec{r}_1\right)$ are the spatial electronic orbitals used to construct the Slater determinant. Since equations (4) are solved for fixed values of the muonic orbital parameters and a certain geometry of clamped nuclei, an extra energy optimization must be done to determine the best energy optimized parameters: $\alpha$, $\vec{R}_c$ and $\vec{R}_1,\dots,\vec{R}_q$. This is done adding $U_s^{eff}$, which does not explicitly depend on the electronic variables, as well as the classic nuclear repulsion to the electronic energy derived from the EHF equations and then optimizing the resulting total energy:

$$E_{total}=E_{EHF-s}+U_s^{eff}+\sum_\beta^q\sum_{\gamma\rangle\beta}^q\frac{Z_\beta Z_\gamma}{\left|\vec{R}_\beta-\vec{R}_\gamma\right|}$$

$$E_{EHF-s}=2\sum_i^{N_e/2}\int d\vec{r}_1\,\psi_i^*\left(\vec{r}_1\right)\hat{h}\left(\vec{r}_1\right)\psi_i\left(\vec{r}_1\right)-2\sum_i^{N_e/2}\int d\vec{r}_1\psi_i^*\left(\vec{r}_1\right)\left(\frac{erf\left[\sqrt{2\alpha}\left|\vec{r}_1-\vec{R}_c\right|\right]}{\left|\vec{r}_1-\vec{R}_c\right|}\right)\psi_i\left(\vec{r}_1\right)$$
$$+\sum_i^{N_e/2}\sum_j^{N_e/2}\int d\vec{r}_1\,\psi_i^*\left(\vec{r}_1\right)\left(2\hat{J}_j\left(\vec{r}_1\right)-\hat{K}_j\left(\vec{r}_1\right)\right)\psi_i\left(\vec{r}_1\right) \qquad (5)$$

In the orthodox electronic structure theory, only the first and third terms are used in the geometry optimization procedure.[1,3] While the s-type gaussian function has been used in our preliminary NEO-HF calculations on the muonic systems,[87-90] more flexible basis sets are desirable for a more reliable description of the muon's vibrations. As an illustrative example the supporting information offers the whole mathematical procedure of deriving the effective potential procedure for [1s1p1d] muonic basis set. As a proof of concept, the resulting equations have been employed computationally in a comparative study of $HCN$ and $\mu CN$ species, the latter results from replacing the proton of the hydrogen cyanide molecule with a muon while in the former the proton is treated as a quantum particle (See Tables S1 and S2 as well as Figures S1-S3 in the supporting information for computational results). In general, the proposed mathematical procedure may be used to transform any given muonic basis set into a *unique* effective potential and based on the known basic integrals in the electronic structure theory,[1,2] an automated algorithm has been constructed to produce an effective potential upon determining the type of gaussian expansion of the muonic orbital (Goli and Shahbazian, under preparation). To derive the general form of the EHF equations an arbitrary expansion of the muonic spatial orbital employing $P$ number of gaussian basis function, $\psi_\mu = \sum_t^P c_t \varphi_t$, is introduced. Incorporating this expansion into equation (2) and after some mathematical manipulations the following set of the EHF equations emerges:

$$\hat{f}^{eff}\left(\vec{r}_1\right)\psi_i\left(\vec{r}_1\right) = \varepsilon_i \psi_i\left(\vec{r}_1\right) \qquad i = 1,\ldots,N_e/2$$

$$\hat{f}^{eff}\left(\vec{r}_1\right) = \hat{h}\left(\vec{r}_1\right) + V_1^{eff}\left(\vec{r}_1\right) + \sum_j^{N_e/2}\left[2\hat{J}_j\left(\vec{r}_1\right) - \hat{K}_j\left(\vec{r}_1\right)\right]$$

$$V_e^{eff} = \sum_i^{N_e} V_i^{eff}\left(\vec{r}_i\right), \qquad V_i^{eff}\left(\vec{r}_i\right) = \sum_{t,w}^{P} c_{tw} v_{tw}\left(\vec{r}_i\right), \qquad v_{tw}\left(\vec{r}_i\right) = -\int d\vec{r}_\mu \frac{\varphi_t^*\left(\vec{r}_\mu\right)\varphi_w\left(\vec{r}_\mu\right)}{\left|\vec{r}_\mu - \vec{r}_i\right|}$$

$$E_{total} = E_{EHF} + U^{eff} + \sum_{\beta}^{q}\sum_{\gamma\rangle\beta}^{q}\frac{Z_\beta Z_\gamma}{\left|\vec{R}_\beta - \vec{R}_\gamma\right|}, \qquad U^{eff} = \sum_{t,w}^{P} c_{tw}\left(T_{tw} + \sum_{\beta}^{q} u_{tw}\left(\vec{R}_\beta\right)\right)$$

$$T_{tw} = \left(-1/2m_\mu\right)\int d\vec{r}_\mu \varphi_t^*\left(\vec{r}_\mu\right)\nabla_\mu^2\varphi_w\left(\vec{r}_\mu\right), \quad u_{tw}^\beta\left(\vec{R}_\beta\right) = Z_\beta \int d\vec{r}_\mu \frac{\varphi_t^*\left(\vec{r}_\mu\right)\varphi_w\left(\vec{r}_\mu\right)}{\left|\vec{R}_\beta - \vec{r}_\mu\right|} \quad (6)$$

These equations are the most general form of the EHF equations and are solved using the computational procedure described in the supporting information. In the present form, the muonic orbital parameters do not appear explicitly in the effective potential however, after the explicit integration of $v_{tw}\left(\vec{r}_i\right)$, $T_{tw}$ and $u_{tw}^\beta\left(\vec{R}_\beta\right)$, using the given basis set, the explicit form of the effective potential is derived.

## III. Computational details

While it is possible to optimize the exponents of the gaussian functions in the SCF cycles during the solution of the EHF equations for each muonic species, this is a time consuming numerical procedure that seems unnecessary since as will be demonstrated, the optimized exponents of the muonic basis sets are not very sensitive to the muon's chemical environment. Thus, a systematic study was done to derive energy-optimized exponents for both muonic basis sets and corresponding electronic basis sets, used to describe electrons surrounding the muon. In order to compare with previous studies on basis set design for the hydrogen isotopes, i.e. proton (H), deuterium (D) and tritium (T),[14,95-102] the exponent optimization procedure was done not only for the muonic species but also representative sets of the hydrides of the second and third row of the periodic table. In total, four molecular sets including fifty-six species: LiX, $BeX_2$, $BX_3$, $CX_4$, $NX_3$, $OX_2$, FX, NaX, $MgX_2$, $AlX_3$, $SiX_4$, $PX_3$, $SX_2$, ClX (X=$\mu$, H, D, T) were considered and since these sets comprise species with multiple quantum nuclei, instead of the EHF equations, the conventional NEO-HF equations were used for the exponent optimization.

The details of the computational protocol are similar to those disclosed in the supporting information for $HCN$ and $\mu CN$ species while the used basis sets are 6-311+g(d) electronic basis set for the central clamped nuclei,[105-107] [1s] and [2s2p2d] for nuclear orbitals and [4s1p] for the electronic basis corresponding to the quantum nuclei; the total basis sets are hereafter denoted as [6-311+g(d)/4s1p:1s] and [6-311+g(d)/4s1p:2s2p2d]. For each quantum nucleus in a species a joint center, a *banquet* atom, was employed for the nuclear and the corresponding electronic basis functions. The exponents of all basis functions corresponding to the quantum nuclei were fully optimized except the exponents of [2s2p2d] basis set, which were optimized only for the muonic species whereas for the hydrogen isotopes, those derived by Hammes-Schiffer and coworkers,[14,95] are used without further optimization (See Table I for the numerical values). In all calculations, the usual point group symmetries of these species were imposed conceiving the banquet atoms as pseudo centers and because of these geometrical symmetries, the nuclear and electronic exponents derived for all centers are the same. The used masses for muon, proton, deuterium and tritium as quantum nuclei in atomic units are $206.768$, $1836$, 3670 and 5496, respectively. Throughout the paper and in tables and figures all inter-nuclear or mean distances have been offered in angstroms while energies are always in atomic units.

The designed [6-311++g(d,p)/4s1p:1s] and [6-311++g(d,p)/4s1p:2s2p2d] basis sets with averaged exponents then was used in conjunction with the EHF equations for ab initio calculations. The fact that equations (4) and (S4) (in supporting information) have been developed for a single heavy particle means that the considered species must contain just a single quantum nucleus. Accordingly, the introduced fifty-six species were reused for the EHF calculations but instead of all, just one of the X nuclei in each species was treated as quantum nucleus, and the others, like the central nucleus, were assumed to be clamped point charges. In order to have a reference for comparison, the standard electronic aug-cc-pVTZ basis set was also used in the EHF calculations, both for the clamped and the quantum nuclei, in the form of [aug-cc-pVTZ:1s] and [aug-cc-pVTZ:2s2p2d] basis sets.[108,109] This electronic basis set contains many types of basis functions covering a large domain of exponent values, some comparable in magnitude to the optimized electronic exponents. Note that in the EHF calculations instead of the distance between the banquet center and the central clamped nucleus (denoted as C) the computed quantum mechanical mean inter-nuclear distances, $\left\langle \hat{r}_{XC} \right\rangle$, have been reported as a more reliable measure of distance in quantum world. Other details of the computational procedure are similar to those disclosed in the supporting information for $HCN$ and $\mu CN$ species. In this paper the relative computational cost of the NEO-HF and the EHF calculations are not considered and we leave a comprehensive discussion on this topic as well as details of performing required integration and corresponding algorithms into a future study.

## IV. Results and discussion

### A. The optimized exponents of the muonic basis sets

Let us first consider the results gained from [6-311+g(d)/4s1p:1s] basis set; the optimized nuclear and electronic exponents have been gathered in Tables S3 and S4 in the supporting information, respectively, while Figures 1 and 2 graphically offer the optimized nuclear exponents for all the quantum nuclei and the optimized electronic exponents for the muonic species, respectively.

Insert Figures 1 and 2

Even a glance at Figure 1 reveals that the nuclear exponents are mainly determined by the mass of the nuclei and the chemical environment, i.e. central atom, has a marginal role. The massive nuclei tend to have larger exponents revealing their tendency for localization and the secondary trends induced by the chemical environment, seem to be similar in the four molecular sets and relatively independent from the mass of the quantum nuclei. Accordingly, in each row of the periodic table the exponents first grow, scanning the central atoms from the left to the right-hand side of the periodic table, reaching to a maximum in group IV and then diminishing. Moreover, the nuclear exponents for species from the third row of the periodic table are always smaller than their counterparts from the second row. While these variations may probably be rationalized based on the effective force constant experienced by the quantum particles in each species, taking the fact that the electron-nucleus lacks in the NEO-HF equations, we prefer not to deduce a physical picture from the optimized exponents. Also, the exponents of $BX_3$ $CX_4$, and $NX_3$ are clearly larger than those of the other species and this is to some extent annoying since in most of the previous studies these three species have been used as the main members of the set used to deduce the average exponent for [1s] nuclear basis set of the hydrogen isotopes whereas they do not seem to be so typical. The average values of the exponents, giving an equal weight to all fourteen species in each set, are given in Table I and generally, they are smaller than those derived for proton and its heavier isotopes in the previous studies namely, 24.2,[98] and 23.8,[100] for X=H species and 35.6,[98] and 35.3,[100] for X=D species.

Insert Table I

If the average is calculated incorporating only $BX_3$, $CX_4$, $NX_3$, $OX_2$ and FX species, used usually as the reference set in the previous studies, then the resulting averages are 23.0 for X=H species and 34.2 for X=D species, which are nearer to the previously derived average exponents.[98,100] The remaining minor numerical differences with the previous studies may be attributed to the different electronic basis sets used in the present and previous studies and the fact that the centers of the electronic and nuclear basis sets have been varied independently in the previous studies.[98,100] By the way, the very idea of an optimal average exponent is to some extent arbitrary and depends on the sort of intended application, thus, the data released in Table S3 may be used to introduce other types of average exponents.

To have a more quantitative view on the mass dependence of the optimized nuclear exponents, the exponent of the s-type gaussian function is re-expressed based on the parameters of the harmonic oscillator model taking into account the fact that the s-type gaussian function is also the ground eigenfunction of the 3D isotropic harmonic oscillator.[92] One easily drives: $\frac{\alpha_X}{\alpha_Y} = \sqrt{\frac{m_X}{m_Y}} \cdot \sqrt{\frac{k_X}{k_Y}}$ (X, Y = $\mu$, H, D, T), where $m$ and $k$ are the mass and the effective force constant experienced by the quantum nucleus, respectively, and since the nature of bonds to the central atoms does not change upon isotope substitution, the first term seems to dominate the variation of the exponents, $\alpha_X \approx \sqrt{\frac{m_X}{m_Y}}\ \alpha_Y$. This relation is numerically in accord with the data in Table S3 demonstrating that knowing just the exponent of one of the isotopically substituted species suffices to fairly well predict the exponents of the same species with other isotopic constitutions. However, some small but systematic deviations are observable from this simple relation stemming from the fact that the effective force constant experienced by a heavier isotope is larger than a lighter isotope. Practically, this simple relation offers the opportunity to develop a general [1s] basis set useful for ab initio studies on systems containing a quantum particle with a variable mass,[88] knowing just the optimized exponent for a single mass value.

In most of previous ab initio studies the used electronic basis sets, for electrons around the quantum nuclei, were those designed originally within the adiabatic paradigm using the point clamped charges as nuclei; while this may be a proper choice for the hydrogen isotopes, it is suspicious to be legitimate in the case of the muon. The averages of the optimized exponents of [4s1p] basis set, given in Table II, confirms this suspicion revealing that while the averaged exponents for species containing the three hydrogen isotopes in a one-to-one comparison are almost the same, the averages for the muonic molecules are distinct from those of the hydrogen isotopes.

Insert Table II

Evidently, since the muon is much lighter than the hydrogen isotopes, it is less capable of localizing electrons around itself resulting to smaller average exponents for all the basis functions. This means that if one tries to use conventional basis sets, designed for the clamped hydrogen nucleus, also for the muon, then more or less only the diffuse functions will be the effective basis functions used to describe electrons around the muon.

The general trends observed for [6-311+g(d)/4s1p:1s] exponents remain almost the same for [6-311+g(d)/4s1p:2s2p2d] basis thus, only the main points are briefly considered while Tables S5 and S6 in the supporting information gather all the optimized muonic and electronic exponents. Figure 3 offers the optimized nuclear exponents for all the muonic species while Table I offers the average values, which may be compared with those derived by Hammes-Schiffer and coworkers in the case of hydrogen isotopes.[14,95]

Insert Figure 3

As is evident, the average exponents of the muonic basis functions are much smaller than those of the hydrogen isotopes and their values conform to that of the average [1s] exponent. Table II displays the averages of the optimized electronic basis set and the averages are near to those derived for [6-311+g(d)/4s1p:1s] basis set revealing the desirable insensitivity of the electronic exponents to the used nuclear basis set.

## B. Comparison of the optimized and averaged basis sets

In order to have a clear picture of the quality of the basis sets constructed from the averaged exponents, in this subsection some results derived from ab initio NEO-HF calculations are compared to the basis sets composed of the optimized exponents. The total energies, the distances and angles between the banquet centers and the central clamped nuclei have been gathered in Tables S7, S8 and S9 in the supporting information, while Figures 4 and 5 offer the differences between the results derived from the optimized and the averaged basis sets graphically.

Insert Figures 4 and 5

A glance at Figure 4a demonstrates that the difference in total energies using the optimized and averaged [6-311+g(d)/4s1p:1s] and [6-311+g(d)/4s1p:2s2p2d] basis sets, except from a single case, is always less than a milli-Hartree. If one compares this sub milli-Hartree difference with the energy differences in Figure 4b between [6-311+g(d)/4s1p:1s] and [6-311+g(d)/4s1p:2s2p2d] basis sets, composed of the optimized exponents, then it is clear that in most cases the energy differences between the optimized basis sets are larger than what is observed in Figure 4a. To have another reference of comparison Figure 5a depicts the difference between the distances of the banquet centers and the central clamped nuclei computed with the averaged and the optimized basis sets revealing that except from a single case, they are less than 0.01 angstroms. Once again if one compares this value to those derived from the optimized basis set in Figure 5b then it is clear that the former is at least an order of magnitude smaller than the latter. Table S9 contains the optimized angles between two banquet centers and the central clamped nucleus in $N\mu_3$, $O\mu_2$, $P\mu_3$ and $S\mu_2$ species and the difference between angles computed with the optimized and averaged basis sets are always less than a degree. Evidently, all these observations point to the fact that the results gained from the averaged basis sets to a large extent replicate those derived from the corresponding optimized basis sets. Even more, these results suggest that if there is a demand for a muonic basis set with a larger quality, it is more economic to add extra functions and construct a larger basis set than optimizing the exponents of the smaller basis set, which is computationally a demanding task. Accordingly, for the ab initio EHF calculations considered in the next subsection only the averaged basis sets are used.

## C. The EHF ab initio calculations

Tables III and IV offer the computed mean inter-nuclear distances and the total energies, respectively, as typical properties derived from the EHF ab initio calculations.

Insert Tables III and IV

Let us first discuss the general trends observed in the data derived from [6-311++g(d,p)/4s1p:1s] and [6-311++g(d,p)/4s1p:2s2p2d] basis sets, which confirm that the EHF derived results are legitimate. Inspection of Table III reveals that as is expected, in all series of the four isotopically substituted species (X = $\mu$, H, D, T) the mean inter-nuclear distances elongate upon substituting the heavier isotope with the lighter ones, and in the case of the muonic species the longest mean distances are observed. Particularly, the distance difference between X=H and X=$\mu$ congener species is much larger than between any two of the congener species containing the hydrogen isotopes. Also, the difference between results gained from the two different nuclear basis sets, i.e. [1s] and [2s2p2d] for the same species, are small for the species containing the hydrogen isotopes but relatively larger for the muonic species demonstrating the need for flexible muonic basis set for ab initio calculations. This is understandable since the muon, because of its smaller mass relative to the hydrogen isotopes, is more prone to anharmonic vibrations and [2s2p2d] muonic basis set accounts for this anharmonicity properly. Inspection of Table IV also reveals the expected trend namely, in the four isotopically substituted congener species the species with the heavier isotope has a more negative total energy relative to that of the species containing the lighter isotope. Evidently, a heavier isotope has a smaller kinetic energy and a larger capacity of accumulating electrons around itself both helping to make the total energy more negative.[87] Also, similar to the case of the mean distances, the difference of the total energies derived from the two different nuclear basis sets for the same species are small for the species containing the hydrogen isotopes but relatively larger for the muonic species. Let us now compare these results with those derived from [aug-cc-pVTZ:1s] and [aug-cc-pVTZ:2s2p2d] basis sets.

Even a glance at Tables III and IV reveals that all trends described for ab initio results gained from [6-311++g(d,p)/4s1p:1s] and [6-311++g(d,p)/4s1p:2s2p2d] basis sets for the muonic species are exactly reproduced from the EHF calculations using much larger [aug-cc-pVTZ:1s] and [aug-cc-pVTZ:2s2p2d] basis sets. This is an independent piece of evidence demonstrating that the designed [4s1p] electronic basis sets are reliable enough to be used for ab initio calculations. To have a more detailed comparison, Figure 6 compares the mean distance and the total energy differences between these two sets of basis sets.

Insert Figure 6

It is clear from Figure 6a that the difference of the mean distances derived from the two basis sets with different electronic but the same muonic basis set is no more than 0.006 Å and in most cases this difference is independent from the used muonic basis set while Figure 6b reveals that the same difference for total energies is usually less than 10 milli-Hartrees.

As a final check the ab initio results gained from [6-311++g(d,p)/4s1p:1s] and [6-311++g(d,p)/4s1p:2s2p2d] basis sets are compared to that derived at the HF/6-311g++(d,p) level for the congener species where all nuclei are treated as clamped point charges. Figure 7 depicts the results for [6-311++g(d,p)/4s1p:2s2p2d] basis set while those derived from [6-311++g(d,p)/4s1p:1s] have been depicted in Figure S4 in the supporting information.

Insert Figure 7

As one expects the computed ab initio results for species containing heavier isotopes are nearer to those derived at the clamped nuclei limit while those of the muonic species are clearly the furthest, once again demonstrating the need to treat muons as quantum particles.

## V. Conclusion

Present study was the first step to exclusively design and simplify the NEO theory for the muonic molecules assuming the muon as a quantum particle and the proposed EHF equations derived for this purpose are simplified version of the coupled NEO-HF equations previously used for ab initio calculations on the muonic species.[87-90] The main idea namely, transforming the muon into an effective potential in the EHF, is quite general and in principle can be extended for any given muonic basis set. However, from a computational perspective it seems that a combination of a number of s-, p-, and d-type Cartesian Gaussian basis functions suffices to construct a flexible enough, but not too complicated, effective potential. The introduction of the general concept of the effective potential also opens the door for "semi-empirical" design of these potentials for quantitative reproduction of certain properties of the muonic systems like the hyperfine coupling constants, which are directly linked to the muon spin resonance spectroscopy.[57,59] Though our target in this and future studies are the muonic molecules however, the developed formalism maybe employed as well for molecular systems where just a single proton, deuterium, tritium or even a hypothetical (with an arbitrary mass) positively charged particle must be treated as a quantum particle. The design of the muon-specific basis sets is also an inevitable part of any computational procedure that aims to apply the effective

NEO theory into the muonic molecules. Indeed, careful optimization of the exponents for both muonic and associated electronic basis functions demonstrated that the best energy-optimized exponents are clearly distinct from those derived previously for the hydrogen isotopes or the clamped hydrogen nucleus. Accordingly, it seems legitimate to try the same procedure in future studies not only to design more extended energy-optimized basis sets but also for basis sets that are used to compute the hyperfine coupling constants.

The developed effective formalism is the best that could be done at the NEO-HF level however, for accurate computational studies as also stressed in the introduction, electron-muon and electron-electron correlations must be incorporated into the effective theory. Accordingly, the next phase of development beyond the EHF theory includes extending the formalism within the context of the NEO-DFT,[20-25] which takes care of both electron-muon and electron-electron correlations simultaneously (for a very recent promising proposal within the context of the NEO-DFT see Ref. 110 and 111). This step is particularly important since it promises an accurate reproduction of the vibrational frequency of the muon in the muonic molecules as well as accurate description of subtle electronic variations induced upon the substitution of proton with muon. All these developments are now under consideration in our lab and the results will be offered in a series of reports.

## Conflicts of interest

There are no conflicts of interest to declare.

## Acknowledgments

The authors are grateful to Masumeh Gharabaghi for her detailed reading of a previous draft of this paper and helpful suggestions.

## Tables:

Table I- The average values of the nuclear exponents of [1s] and [2s2p2d] basis sets. *

| | **s** | | **s** | **s** | **p** | **p** | **d** | **d** |
|---|---|---|---|---|---|---|---|---|
| **T** | 40.58 | **T** | *47.80* | *33.73* | *42.87* | *28.18* | *44.84* | *33.82* |
| **D** | 32.41 | **D** | *42.33* | *30.60* | *40.26* | *29.50* | *36.89* | *25.51* |
| **H** | 21.84 | **H** | *29.29* | *20.18* | *26.17* | *18.22* | *23.71* | *16.73* |
| $\mu$ | 5.75 | $\mu$ | 8.27 | 6.71 | 6.00 | 4.19 | 6.66 | 4.59 |

*The [2s2p2d] exponents of the hydrogen isotopes are those derived previously by Hammes-Schiffer and coworkers (Ref. 14 and 95) and have been presented here only for comparison.

Table II- The average values of the electronic exponents of [4s1p:1s] and [4s1p:2s2p2d] basis sets.

| ***1s*** | | | | | |
|---|---|---|---|---|---|
| | **s** | **s** | **s** | **s** | **p** |
| **T** | 10.87 | 2.16 | 0.55 | 0.16 | 0.65 |
| **D** | 10.00 | 2.06 | 0.54 | 0.16 | 0.64 |
| **H** | 8.49 | 1.88 | 0.51 | 0.16 | 0.63 |
| $\mu$ | 4.21 | 1.20 | 0.37 | 0.12 | 0.58 |
| ***2s2p2d*** | | | | | |
| | **s** | **s** | **s** | **s** | **p** |
| **T** | 10.72 | 2.16 | 0.55 | 0.16 | 0.54 |
| **D** | 9.94 | 2.09 | 0.55 | 0.16 | 0.47 |
| **H** | 8.23 | 1.88 | 0.52 | 0.16 | 0.46 |
| $\mu$ | 4.22 | 1.23 | 0.39 | 0.12 | 0.47 |

Table III- The computed mean inter-nuclear distances between the quantum nucleus and the central nucleus offered in Angstroms. The clamped nuclei are all at the left-hand side of the symbol “:” while the quantum nucleus is at the right-hand side.

| | 6-311++g(d,p)/4s1p | | aug-cc-pVTZ | | | 6-311++g(d,p)/4s1p | | aug-cc-pVTZ | |
|---|---|---|---|---|---|---|---|---|---|
| | **1s** | **2s2p2d** | **1s** | **2s2p2d** | | **1s** | **2s2p2d** | **1s** | **2s2p2d** |
| **Li:T** | 1.618 | 1.619 | 1.623 | 1.624 | **Na:T** | 1.921 | 1.922 | 1.932 | 1.934 |
| **Li:D** | 1.621 | 1.623 | 1.627 | 1.628 | **Na:D** | 1.925 | 1.925 | 1.936 | 1.937 |
| **Li:H** | 1.630 | 1.632 | 1.635 | 1.637 | **Na:H** | 1.933 | 1.934 | 1.942 | 1.945 |
| **Li:μ** | 1.688 | 1.693 | 1.692 | 1.695 | **Na:μ** | 1.986 | 1.993 | 1.992 | 1.998 |
| | | | | | | | | | |
| **BeH:T** | 1.348 | 1.348 | 1.348 | 1.347 | **MgH:T** | 1.724 | 1.724 | 1.724 | 1.724 |
| **BeH:D** | 1.351 | 1.351 | 1.352 | 1.351 | **MgH:D** | 1.727 | 1.728 | 1.727 | 1.727 |
| **BeH:H** | 1.360 | 1.359 | 1.360 | 1.359 | **MgH:H** | 1.735 | 1.736 | 1.735 | 1.735 |
| **BeH:μ** | 1.415 | 1.413 | 1.417 | 1.413 | **MgH:μ** | 1.788 | 1.790 | 1.788 | 1.790 |
| | | | | | | | | | |
| **$BH_2$:T** | 1.206 | 1.205 | 1.204 | 1.202 | **$AlH_2$:T** | 1.594 | 1.595 | 1.597 | 1.597 |
| **$BH_2$:D** | 1.210 | 1.208 | 1.208 | 1.206 | **$AlH_2$:D** | 1.597 | 1.599 | 1.601 | 1.600 |
| **$BH_2$:H** | 1.218 | 1.216 | 1.216 | 1.213 | **AlH2:H** | 1.606 | 1.607 | 1.609 | 1.608 |
| **$BH_2$:μ** | 1.273 | 1.265 | 1.272 | 1.264 | **AlH2:μ** | 1.662 | 1.661 | 1.665 | 1.664 |
| | | | | | | | | | |
| **$CH_3$:T** | 1.101 | 1.098 | 1.098 | 1.096 | **$SiH_3$:T** | 1.493 | 1.494 | 1.494 | 1.494 |
| **$CH_3$:D** | 1.104 | 1.102 | 1.101 | 1.099 | **$SiH_3$:D** | 1.497 | 1.498 | 1.498 | 1.497 |
| **$CH_3$:H** | 1.112 | 1.108 | 1.109 | 1.106 | **$SiH_3$:H** | 1.505 | 1.505 | 1.507 | 1.505 |
| **$CH_3$:μ** | 1.163 | 1.152 | 1.162 | 1.151 | **$SiH_3$:μ** | 1.560 | 1.557 | 1.563 | 1.559 |
| | | | | | | | | | |
| **$NH_2$:T** | 1.015 | 1.013 | 1.013 | 1.011 | **$PH_2$:T** | 1.424 | 1.424 | 1.424 | 1.424 |
| **$NH_2$:D** | 1.018 | 1.016 | 1.016 | 1.014 | **$PH_2$:D** | 1.428 | 1.428 | 1.428 | 1.427 |
| **$NH_2$:H** | 1.025 | 1.021 | 1.023 | 1.020 | **$PH_2$:H** | 1.436 | 1.436 | 1.437 | 1.435 |
| **$NH_2$:μ** | 1.073 | 1.061 | 1.072 | 1.059 | **$PH_2$:μ** | 1.490 | 1.487 | 1.492 | 1.488 |
| | | | | | | | | | |
| **OH:T** | 0.955 | 0.953 | 0.955 | 0.953 | **SH:T** | 1.347 | 1.346 | 1.345 | 1.345 |
| **OH:D** | 0.958 | 0.955 | 0.958 | 0.955 | **SH:D** | 1.350 | 1.350 | 1.349 | 1.348 |
| **OH:H** | 0.964 | 0.961 | 0.965 | 0.961 | **SH:H** | 1.358 | 1.358 | 1.357 | 1.356 |
| **OH:μ** | 1.010 | 0.999 | 1.011 | 0.997 | **SH:μ** | 1.409 | 1.405 | 1.409 | 1.406 |
| | | | | | | | | | |
| **F:T** | 0.910 | 0.910 | 0.912 | 0.910 | **Cl:T** | 1.285 | 1.285 | 1.282 | 1.282 |
| **F:D** | 0.913 | 0.913 | 0.915 | 0.913 | **Cl:D** | 1.289 | 1.289 | 1.286 | 1.285 |
| **F:H** | 0.920 | 0.919 | 0.922 | 0.918 | **Cl:H** | 1.296 | 1.296 | 1.293 | 1.293 |
| **F:μ** | 0.966 | 0.958 | 0.967 | 0.953 | **Cl:μ** | 1.344 | 1.341 | 1.344 | 1.341 |

Table IV- The computed total energies offered in Hartrees. The clamped nuclei are all at the left-hand side of the symbol ":" while the quantum nucleus is at the right-hand side.

| | **6-311++g(d,p)/4s1p** | | **aug-cc-pVTZ** | | | **6-311++g(d,p)/4s1p** | | **aug-cc-pVTZ** | |
|---|---|---|---|---|---|---|---|---|---|
| | **1s** | **2s2p2d** | **1s** | **2s2p2d** | | **1s** | **2s2p2d** | **1s** | **2s2p2d** |
| **Li:T** | -7.9633 | -7.9633 | -7.9647 | -7.9647 | **Na:T** | -162.3563 | -162.3564 | -162.3696 | -162.3696 |
| **Li:D** | -7.9587 | -7.9588 | -7.9601 | -7.9601 | **Na:D** | -162.3520 | -162.3521 | -162.3651 | -162.3652 |
| **Li:H** | -7.9489 | -7.9490 | -7.9499 | -7.9500 | **Na:H** | -162.3425 | -162.3427 | -162.3553 | -162.3554 |
| **Li:μ** | -7.8916 | -7.8919 | -7.8915 | -7.8917 | **Na:μ** | -162.2875 | -162.2880 | -162.2987 | -162.2992 |
| **BeH:T** | -15.7469 | -15.7470 | -15.7481 | -15.7481 | **MgH:T** | -200.7081 | -200.7082 | -200.7156 | -200.7156 |
| **BeH:D** | -15.7420 | -15.7420 | -15.7431 | -15.7431 | **MgH:D** | -200.7035 | -200.7036 | -200.7109 | -200.7109 |
| **BeH:H** | -15.7313 | -15.7313 | -15.7321 | -15.7322 | **MgH:H** | -200.6934 | -200.6935 | -200.7005 | -200.7006 |
| **BeH:μ** | -15.6685 | -15.6688 | -15.6685 | -15.6687 | **MgH:μ** | -200.6346 | -200.6349 | -200.6406 | -200.6408 |
| **$BH_2$:T** | -26.3721 | -26.3722 | -26.3751 | -26.3752 | **$AlH_2$:T** | -243.6144 | -243.6146 | -243.6216 | -243.6216 |
| **$BH_2$:D** | -26.3669 | -26.3670 | -26.3698 | -26.3700 | **$AlH_2$:D** | -243.6096 | -243.6097 | -243.6167 | -243.6167 |
| **$BH_2$:H** | -26.3556 | -26.3557 | -26.3582 | -26.3584 | **AlH2:H** | -243.5991 | -243.5992 | -243.6059 | -243.6059 |
| **$BH_2$:μ** | -26.2892 | -26.2899 | -26.2908 | -26.2914 | **AlH2:μ** | -243.5374 | -243.5377 | -243.5431 | -243.5433 |
| **$CH_3$:T** | -40.1834 | -40.1837 | -40.1881 | -40.1883 | **$SiH_3$:T** | -291.2294 | -291.2295 | -291.2373 | -291.2373 |
| **$CH_3$:D** | -40.1782 | -40.1784 | -40.1827 | -40.1830 | **$SiH_3$:D** | -291.2244 | -291.2245 | -291.2322 | -291.2322 |
| **$CH_3$:H** | -40.1667 | -40.1670 | -40.1709 | -40.1713 | **$SiH_3$:H** | -291.2135 | -291.2137 | -291.2211 | -291.2211 |
| **$CH_3$:μ** | -40.0992 | -40.1004 | -40.1023 | -40.1034 | **$SiH_3$:μ** | -291.1501 | -291.1505 | -291.1566 | -291.1569 |
| **$NH_2$:T** | -56.1891 | -56.1897 | -56.1955 | -56.1957 | **$PH_2$:T** | -342.4543 | -342.4545 | -342.4643 | -342.4643 |
| **$NH_2$:D** | -56.1839 | -56.1844 | -56.1901 | -56.1904 | **$PH_2$:D** | -342.4493 | -342.4496 | -342.4592 | -342.4593 |
| **$NH_2$:H** | -56.1724 | -56.1731 | -56.1783 | -56.1788 | **$PH_2$:H** | -342.4385 | -342.4388 | -342.4482 | -342.4483 |
| **$NH_2$:μ** | -56.1052 | -56.1073 | -56.1100 | -56.1117 | **$PH_2$:μ** | -342.3754 | -342.3762 | -342.3843 | -342.3847 |
| **OH:T** | -76.0283 | -76.0293 | -76.0365 | -76.0368 | **SH:T** | -398.6788 | -398.6793 | -398.6905 | -398.6905 |
| **OH:D** | -76.0232 | -76.0241 | -76.0312 | -76.0316 | **SH:D** | -398.6739 | -398.6745 | -398.6855 | -398.6856 |
| **OH:H** | -76.0119 | -76.0130 | -76.0197 | -76.0203 | **SH:H** | -398.6633 | -398.6640 | -398.6746 | -398.6748 |
| **OH:μ** | -75.9457 | -75.9489 | -75.9528 | -75.9552 | **SH:μ** | -398.6018 | -398.6031 | -398.6122 | -398.6128 |
| **F:T** | -100.0286 | -100.0300 | -100.0378 | -100.0382 | **Cl:T** | -460.0729 | -460.0737 | -460.0848 | -460.0849 |
| **F:D** | -100.0236 | -100.0245 | -100.0327 | -100.0332 | **Cl:D** | -460.0682 | -460.0691 | -460.0799 | -460.0801 |
| **F:H** | -100.0126 | -100.0139 | -100.0216 | -100.0224 | **Cl:H** | -460.0580 | -460.0590 | -460.0694 | -460.0696 |
| **F:μ** | -99.9486 | -99.9535 | -99.9575 | -99.9606 | **Cl:μ** | -459.9987 | -460.0009 | -460.0092 | -460.0102 |

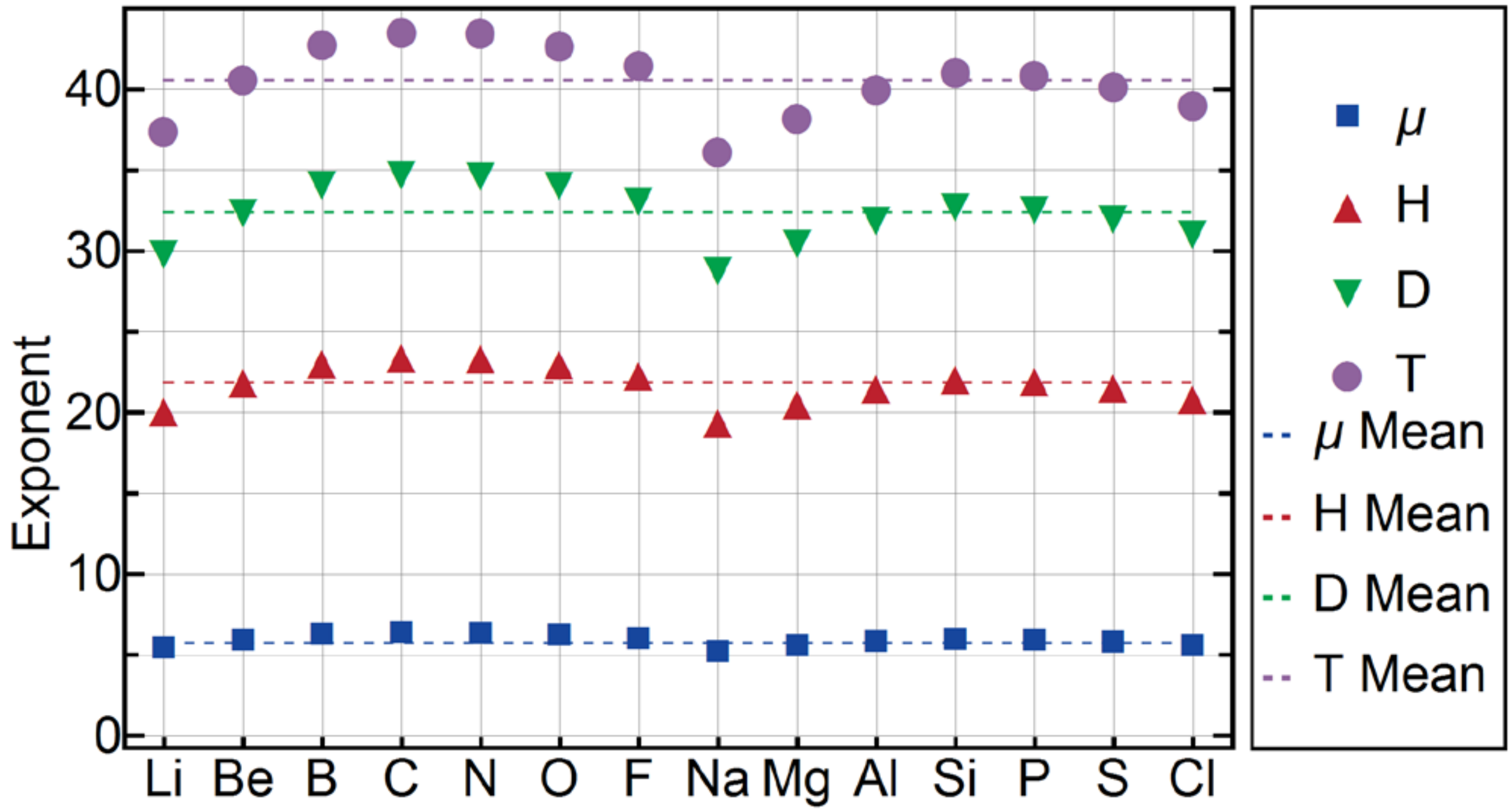

Figure 1- The optimized nuclear exponents of [1s] basis set of all the considered species (see text for details). The average values of the exponents for each set of species with the same isotopic constitution are indicated by dashed lines.

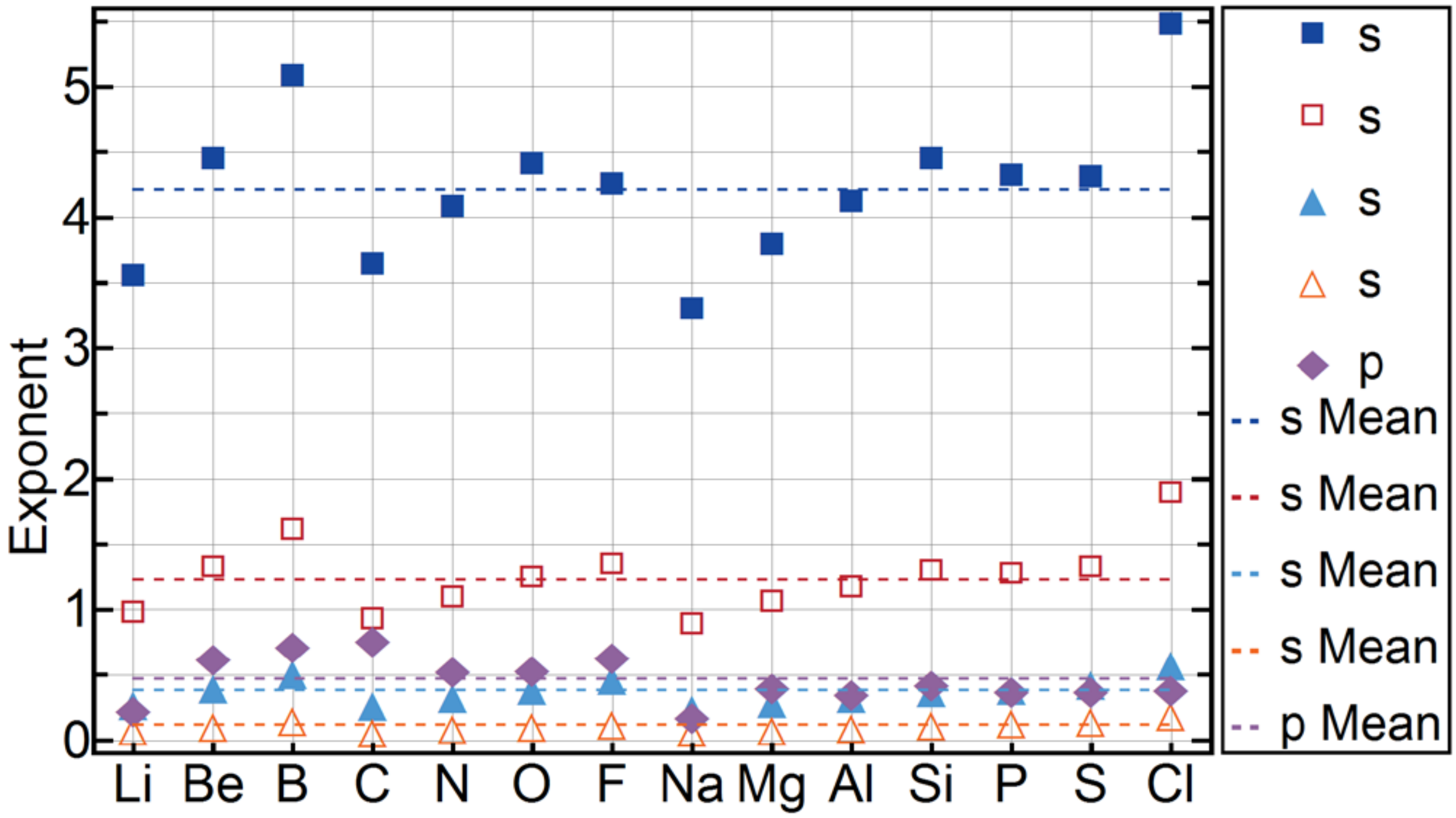

Figure 2- The optimized electronic exponents of [4s1p:1s] basis set of all the considered muonic species (see text for details). The average values of each set of the basis functions are indicated by dashed lines.

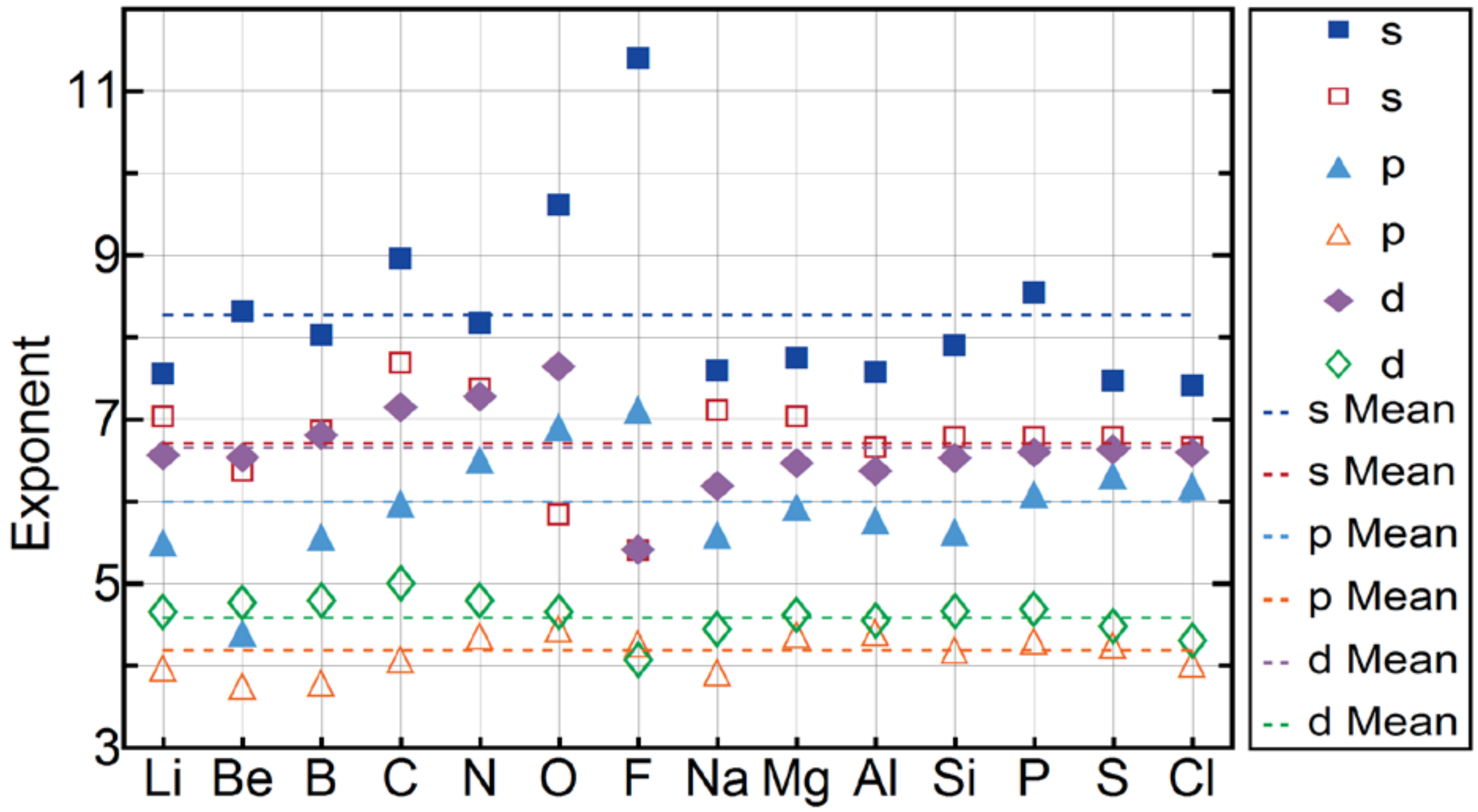

Figure 3- The optimized nuclear exponents of [2s2p2d] basis set of all the considered muonic species (see text for details). The average values of each set of the basis functions are indicated by dashed lines.

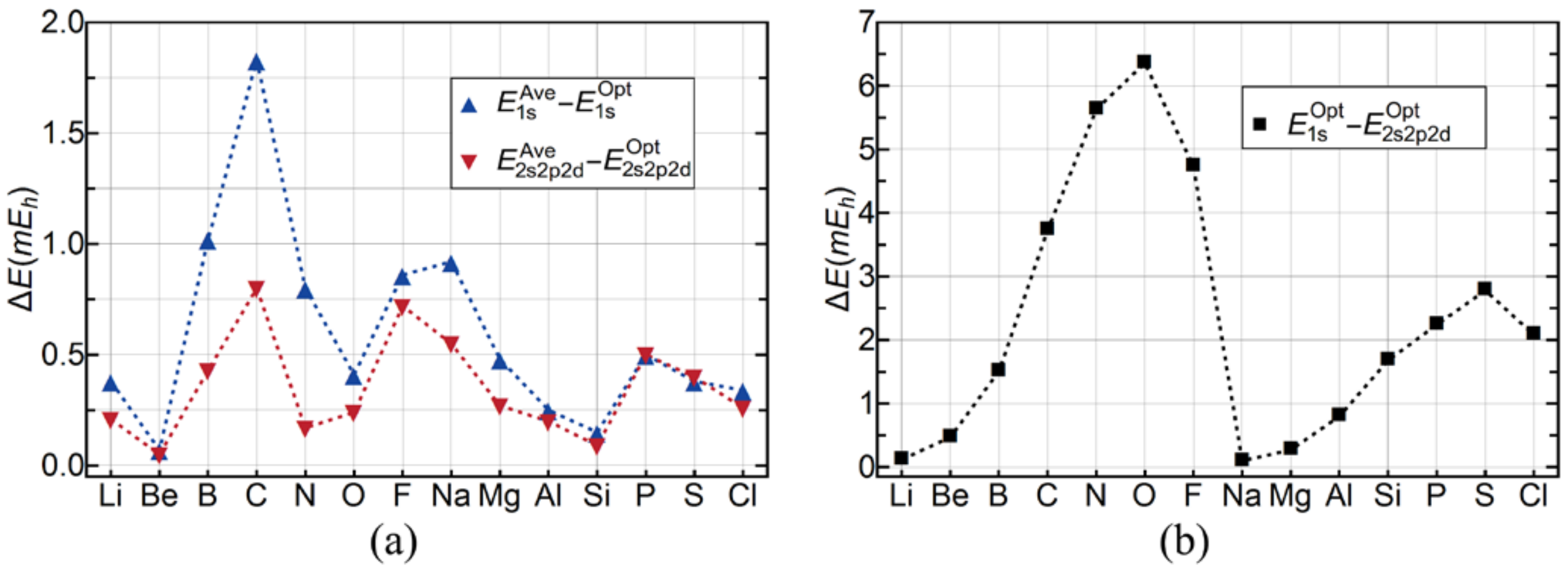

Figure 4- The difference in total energies of all the considered muonic species (see text for details) at NEO-HF level using (a) the optimized and averaged [6-311+g(d)/4s1p:1s] and [6-311+g(d)/4s1p:2s2p2d] basis sets and (b) the optimized [6-311+g(d)/4s1p:1s] and [6-311G+g(d)/4s1p:2s2p2d] basis sets.

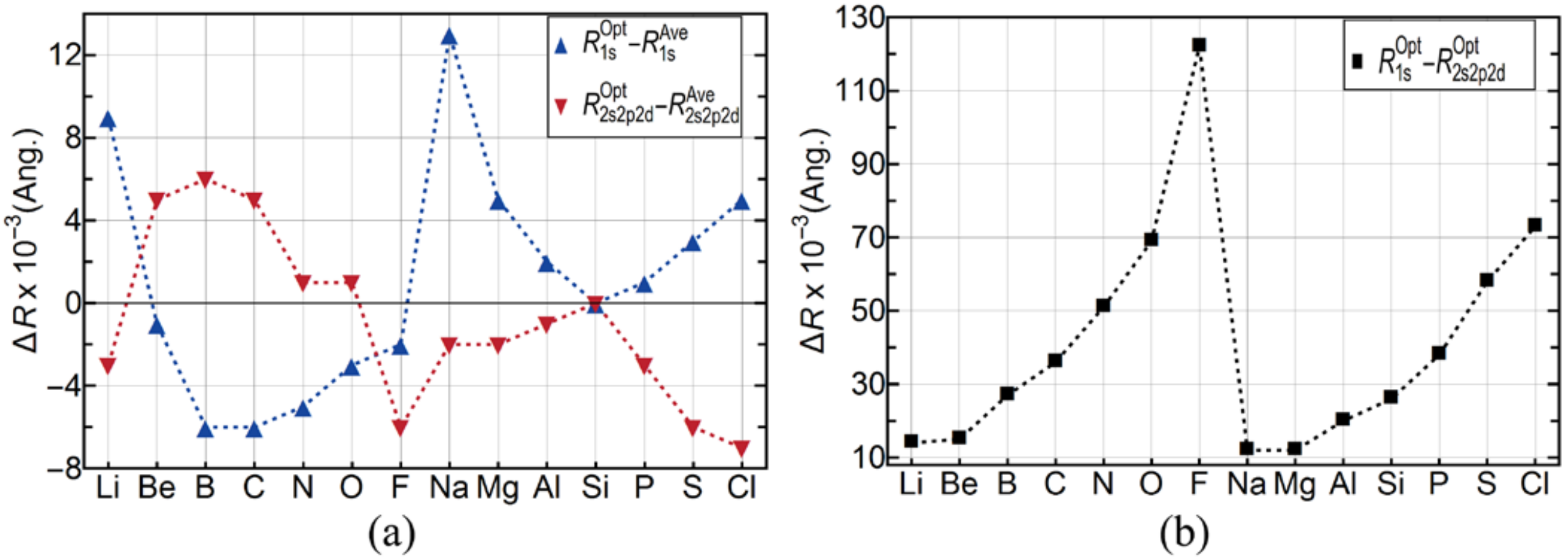

Figure 5- The difference in the distances between the banquet centers and the central atom of all the considered muonic species (see text for details) at NEO-HF level using (a) the optimized and averaged [6-311+g(d)/4s1p:1s] and [6-311G+g(d)/4s1p:2s2p2d] basis sets and (b) the optimized [6-311+g(d)/4s1p:1s] and [6-311G+g(d)/4s1p:2s2p2d] basis sets.

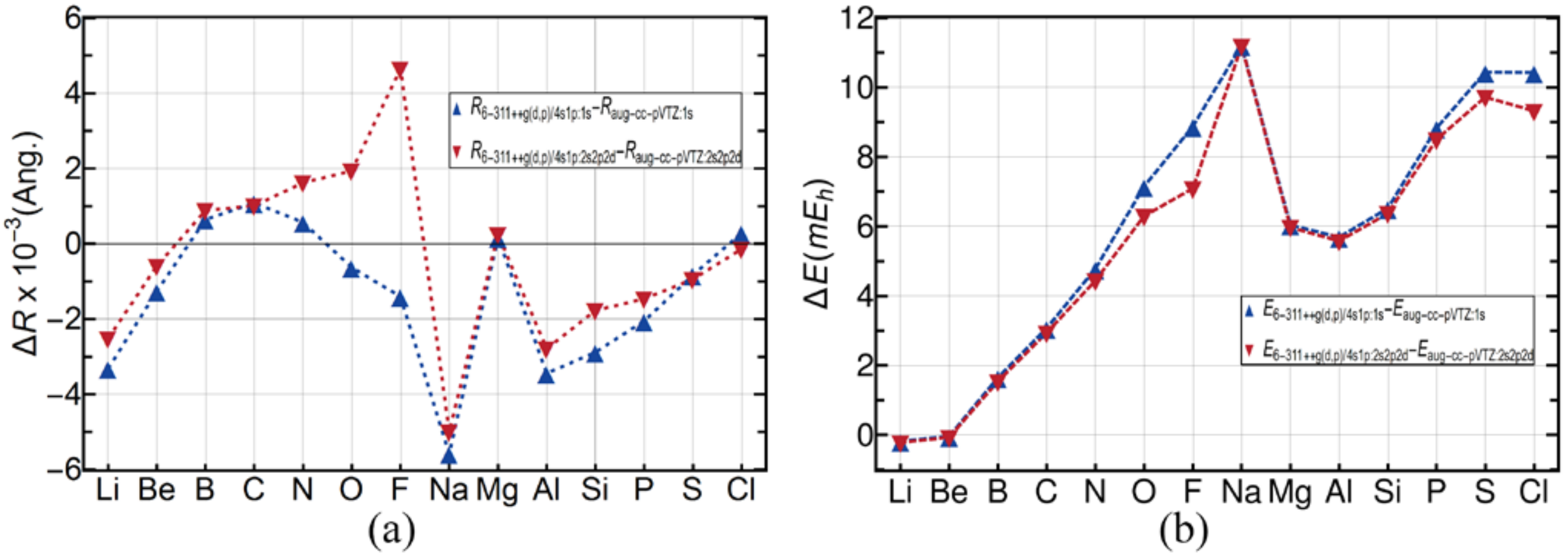

Figure 6- The difference in (a) the mean inter-nuclear distances (between the muon and the central clamped nucleus) in angstroms and (b) total energies (in milli-Hartrees) of all singly-substituted X= $\mu$ species (see text for details) at NEO-HF level using averaged [6-311++g(d,p)/4s1p:1s] and [aug-cc-pVTZ:1s] as well as [6-311++g(d,p)/4s1p:2s2p2d] and [aug-cc-pVTZ:2s2p2d] basis sets.

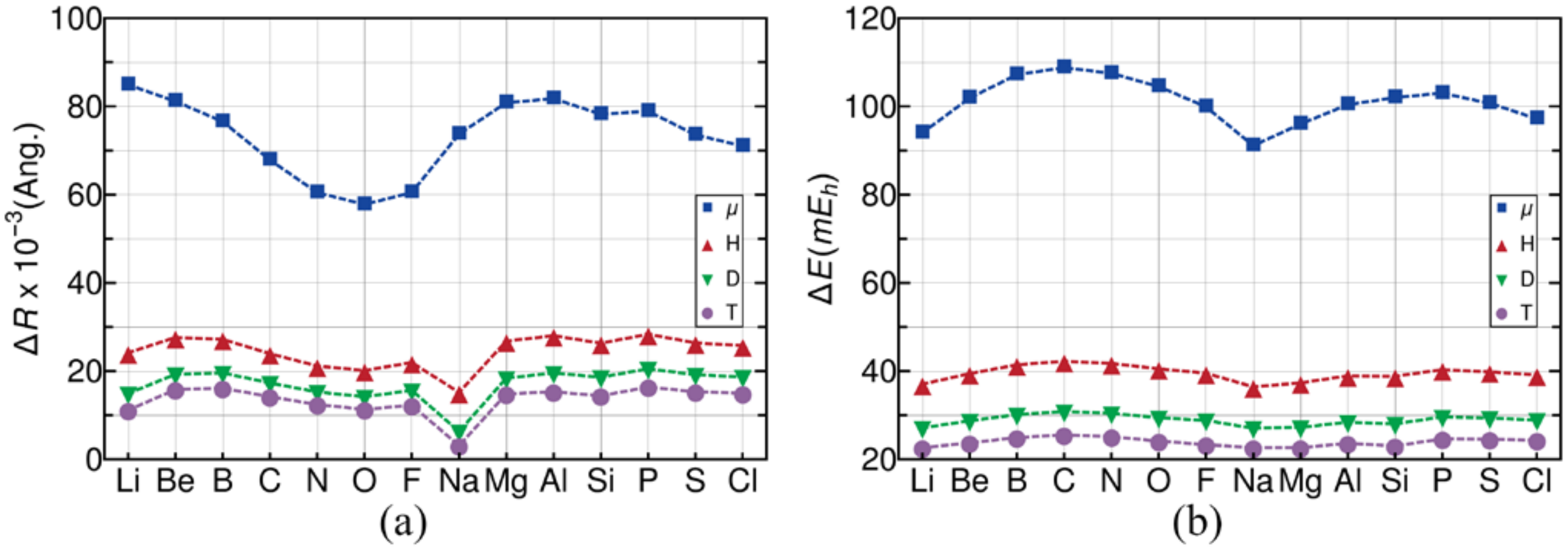

Figure 7- The difference in (a) the mean inter-nuclear distances (between the quantum nucleus and the central atom) in angstroms and (b) the total energies (in milli-Hartrees) of the singly-substituted X= $\mu$, H, D, T species (see text for details) relative to their clamped nucleus counterparts, computed at NEO-HF/[6-311++g(d,p)/4s1p:2s2p2d] and HF/6-311++g(d,p) levels.

# Supporting Information

## Toward a muon-specific electronic structure theory: Effective electronic Hartree-Fock equations for the muonic molecules

Milad Rayka[1], Mohammad Goli[2,*] and Shant Shahbazian[1,*]

[1] *Department of Physics and Department of Physical and Computational Chemistry, Shahid Beheshti University, G. C., Evin, Tehran, Iran, 19839, P.O. Box 19395-4716.*

[2] *School of Nano Science, Institute for Research in Fundamental Sciences (IPM), Tehran 19395-5531, Iran*

E-mails:

Mohammad Goli : mgoli2019@gmail.com

Shant Shahbazian: sh_shahbazian@sbu.ac.ir

[*] Corresponding authors

# Table of contents

## Deriving the EHF equations for [1s1p1d] muonic basis set and their computational implementation

Previous computational experiences reveal that a combination of the s-, p- and d-type Cartesian gaussian functions suffices for a relatively accurate description of the nuclear spatial orbital at the NEO-HF level [S1]. Accordingly, a [1s1p1d] muonic basis set and a [4s1p] electronic basis set are used to expand the muonic spatial orbital and to describe the electronic distribution around the muon, respectively. A joint center, a *banquet* atom, at the z-axis is employed for all the muonic and the electronic basis functions, and the clamped carbon and nitrogen nuclei are placed at the same axis while the center of the coordinate system is fixed at the clamped carbon nucleus. In order to describe the electronic distribution around the clamped nuclei Pople-type 6-311+g(d) basis set is placed at the positions of the clamped nuclei [S2-S4]. For the muonic and corresponding electronic basis functions all parameters, i.e., the SCF linear coefficients, the exponents of the gaussian functions and the position of the joint center of the basis functions are optimized variationally during the NEO-HF calculation. In the process of the optimization of the exponents of the gaussian basis functions, the exponents of each type of gaussian function, e.g., $p$-type, are constrained to be the same for all members of the subset, e.g., $p_x, p_y, p_z$, and are denoted as $\alpha_s, \alpha_p, \alpha_d$ (the $\mu$ subscript is dropped hereafter for brevity). On the other hand, for the electronic basis sets centered on the clamped nuclei only the SCF coefficients are optimized, as is usual in the course of the conventional HF calculations [S5]. The geometry of the clamped nuclei is optimized using the analytical gradients of the total energy [S6], while for the optimization of the exponents of the basis functions a non-gradient optimization algorithm is used as described previously [S7-S10]. The mass of the muon was fixed at $206.768$ in atomic units throughout the calculations and the whole NEO-HF calculations are also redone on hydrogen cyanide molecule where the proton is conceived as a quantum particle with a mass fixed at $1836$ in atomic units.

Table S1 offers the variationally determined exponents and the SCF coefficients of the muonic and the protonic basis functions; from the original ten basis functions in [1s1p1d] basis set, only five basis functions namely, $s, p_z, d_{x^2}, d_{y^2}, d_{z^2}$, have non-zero SCF coefficients.

Table S1- The variationally optimized SCF coefficients and exponents of the muonic and the protonic basis functions derived from the NEO-HF calculations.

| Type of basis functions | $\mu CN$ | | $HCN$ | |
|---|---|---|---|---|
| | **SCF coefficients** | **exponents** | **SCF coefficients** | **exponents** |
| **s** | 0.789 | 7.84 | 0.826 | 27.62 |
| **$p_z$** | -0.206 | 5.51 | -0.218 | 21.86 |
| **$d_x^2$** | 0.143 | 5.94 | 0.104 | 22.73 |
| **$d_y^2$** | 0.143 | | 0.104 | |
| **$d_z^2$** | 0.059 | | 0.059 | |

The normalized muonic and protonic spatial orbitals are both linear combinations of these five basis functions:

$$\psi_{\mu-spd} = c_1\varphi_s + c_2\varphi_{p_z} + c_3\varphi_{d_{x^2}} + c_4\varphi_{d_{y^2}} + c_5\varphi_{d_{z^2}}, \quad \psi_{proton-spd} = c'_1\varphi_s + c'_2\varphi_{p_z} + c'_3\varphi_{d_{x^2}} + c'_4\varphi_{d_{y^2}} + c'_5\varphi_{d_{z^2}}$$

$$\varphi_s = N_s Exp\left(-\alpha_s \left|\vec{r}_\mu - \vec{R}_c\right|^2\right), \quad N_s = \left(\frac{8\alpha_s^3}{\pi^3}\right)^{\frac{1}{4}},$$

$$\varphi_{p_z} = N_p \overline{z}_{\mu c} Exp\left(-\alpha_p \left|\vec{r}_\mu - \vec{R}_c\right|^2\right), \quad N_p = \left(\frac{128\alpha_p^5}{\pi^3}\right)^{1/4}, \quad \overline{z}_{\mu c} = z_\mu - Z_c$$

$$\varphi_{d_{x^2}} = N_d \overline{x}_{\mu c}^2 Exp\left(-\alpha_d \left|\vec{r}_\mu - \vec{R}_c\right|^2\right), \quad N_d = \left(\frac{2048\alpha_d^7}{9\pi^3}\right)^{1/4}, \quad \overline{x}_{\mu c}^2 = \left(x_\mu - X_c\right)^2$$

$$\varphi_{d_{y^2}} = N_d \overline{y}_{\mu c}^2 Exp\left(-\alpha_d \left|\vec{r}_\mu - \vec{R}_c\right|^2\right), \quad \overline{y}_{\mu c}^2 = \left(y_\mu - Y_c\right)^2$$

$$\varphi_{d_{z^2}} = N_d \overline{z}_{\mu c}^2 Exp\left(-\alpha_d \left|\vec{r}_\mu - \vec{R}_c\right|^2\right), \quad \overline{z}_{\mu c}^2 = \left(z_\mu - Z_c\right)^2 \qquad \text{(S1)}$$

Figure S1 compares the one-particle densities, $\rho_\mu = \psi_{\mu-spd}^2$ and $\rho_{proton} = \psi_{proton-spd}^2$, and in line with the numerical data in Table S1 it is clear that the latter is much more concentrated than the former while the anisotropic nature of both distributions is evident from the offered counter maps.

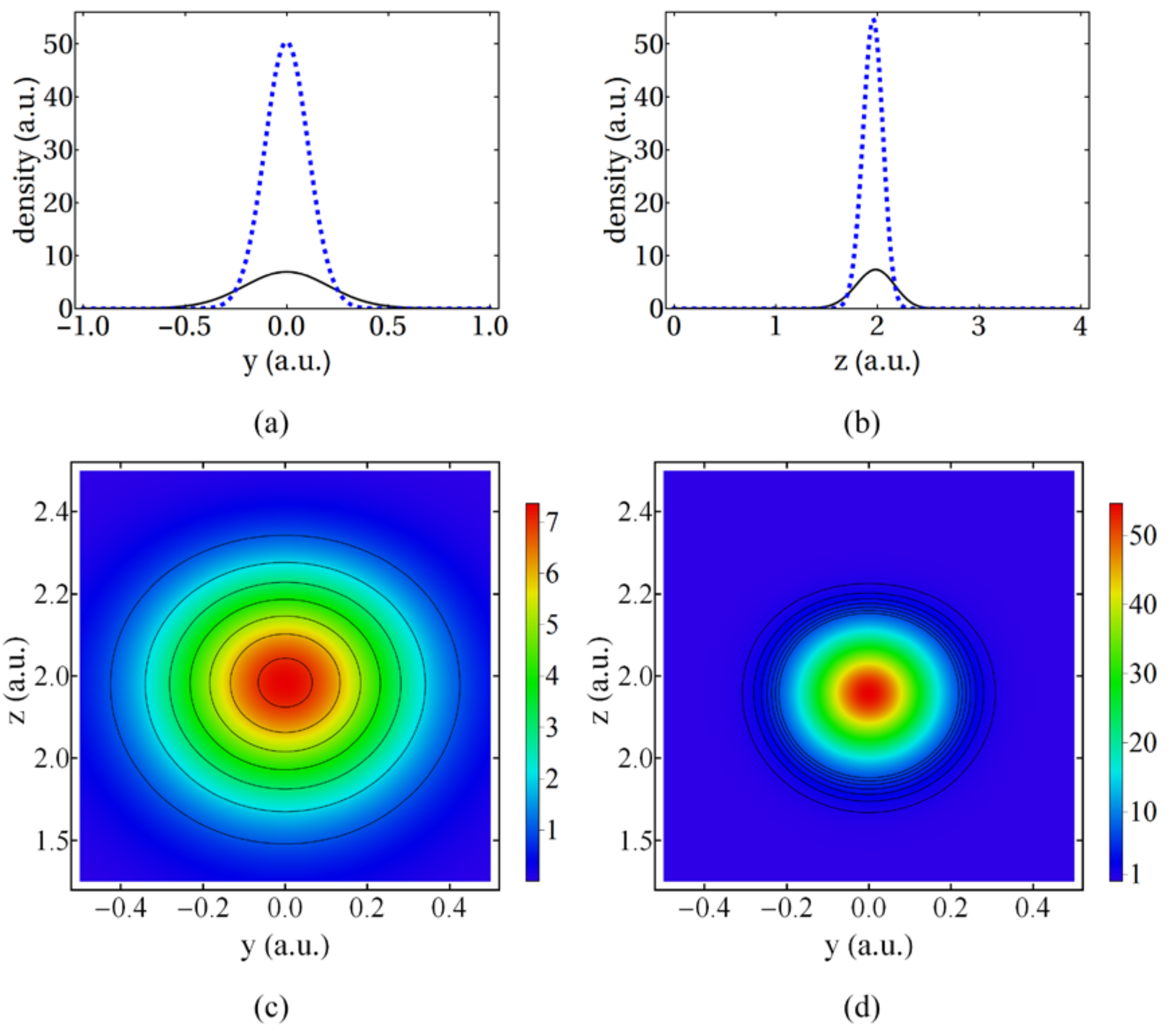

Figure S1- a) The one-particle protonic (dashed line) and muonic (full line) densities depicted along a y-axis, which goes through muon and is perpendicular to the z-axis. b) The same densities along the z-axis. The contour maps of the muonic (c) and the protonic (d) one-particle densities in $\mu CN$ and $HCN$ depicted at yz-plane, respectively (the contours lines are from $\rho$ = 1 to 7, increased in integer steps). The clamped carbon nucleus is placed at the center of coordinate system while the clamped nitrogen nucleus and the banquet atom are placed at the negative and the positive sides of the z-axis, respectively.

Table S2 offers the total, the electronic, and the nuclear kinetic energies as well as the inter-nuclear distances computed at the NEO-HF level (the banquet atom is used as the third center).

Table S2- Some results of the NEO-HF calculations.

| **Energy** | $\mu CN$ | $HCN$ |
|---|---|---|
| **total** | -92.79837 | -92.86175 |
| **electronic kinetic** | 92.72631 | 92.81344 |
| $\mu$ **or proton kinetic** | 0.04216 | 0.01828 |
| **Distances** | | |
| **C-N** | 1.128 | 1.127 |
| **Bq-C** | 1.132 | 1.082 |

The results demonstrate that upon the substitution of the proton with the muon, the latter's mean distribution and the kinetic energy increase relative to those of the former's. Also, the particle with the larger mass, because of its larger localization, is capable of localizing electrons more efficiently [S7], thus the electronic kinetic energy of the hydrogen cyanide molecule is larger than its muonic analog.

Taking into account that the NEO-HF calculation yields the anisotropy and anharmonicity of muon's vibrations using $\psi_{\mu-spd}$, it seems $\psi_{\mu-spd}$ to be a proper model to derive $V^{eff}=V_e^{eff}+U^{eff}$. Incorporating $\psi_{\mu-spd}$ into equation (2), in the main text, and after some mathematical manipulations, the corresponding effective electron-muon interaction, $V_{e-spd}^{eff}$, is derived:

$$V_{e-spd}^{eff}=\sum_i^{N_e}V_{spd}^{eff}\left(\vec{r}_i\right)$$

$$V_{spd}^{eff}\left(\vec{r}_i\right)=-[c_{11}N_{ss}\left(\frac{2\pi}{\alpha_{ss}}\right)F_{0,ss}^i+c_{12}N_{sp}\left(\frac{4\pi}{\alpha_{sp}}\right)\overline{z}_{ic}F_{1,sp}^i+c_{22}N_{pp}\left(\frac{\pi}{\alpha_{pp}^2}\right)(F_{0,pp}^i-F_{1,pp}^i+2\alpha_{pp}\overline{z}_{ic}^2F_{2,pp}^i)$$

$$+N_{sd}\left(\frac{2\pi}{\alpha_{sd}^{\ 2}}\right)\{(c_{13}+c_{14}+c_{15})(F_{0,sd}^i-F_{1,sd}^i)+2\alpha_{sd}(c_{13}\overline{x}_{ic}^2+c_{14}\overline{y}_{ic}^2+c_{15}\overline{z}_{ic}^2)F_{2,sd}^i\}$$

$$+N_{pd}\left(\frac{2\pi}{\alpha_{pd}^2}\right)\overline{z}_{ic}\{(c_{23}+c_{24}+3c_{25})(F_{1,pd}^i-F_{2,pd}^i)+2\alpha_{pd}(c_{23}\overline{x}_{ic}^2+c_{24}\overline{y}_{ic}^2+c_{25}\overline{z}_{ic}^2)F_{3,pd}^i\}$$

$$+N_{dd}\left(\frac{\pi}{2\alpha_{dd}^3}\right)\{3(c_{33}+c_{44}+c_{55})(F_{0,dd}^i-2F_{1,dd}^i+F_{2,dd}^i)+12\alpha_{dd}(c_{33}\overline{x}_{ic}^2+c_{44}\overline{y}_{ic}^2+c_{55}\overline{z}_{ic}^2)$$
$$(F_{2,dd}^i-F_{3,dd}^i)+4\alpha_{dd}^2(c_3\overline{x}_{ic}^2+c_4\overline{y}_{ic}^2+c_5\overline{z}_{ic}^2)^2F_{4,dd}^i+2(c_{34}+c_{35}+c_{45})(F_{0,dd}^i-2F_{1,dd}^i+F_{2,dd}^i)$$
$$+4\alpha_{dd}((c_{34}+c_{35})\overline{x}_{ic}^2+(c_{34}+c_{45})\overline{y}_{ic}^2+(c_{35}+c_{45})\overline{z}_{ic}^2)(F_{2,dd}^i-F_{3,dd}^i)\}] \qquad \text{(S2)}$$

In this expression $c_{tw}=c_tc_w,\ t,w=1-5$ and $\alpha_{kl}=\alpha_k+\alpha_l,\ N_{kl}=N_kN_l,\ k,l=s,p,d$ while $\overline{x}_{ic}^n=\left(x_i-X_c\right)^n,\ n=0-4$ (similarly for $\overline{y}_{ic}^n$ and $\overline{z}_{ic}^n$), and $F_{n,kl}^i=\int_0^1 dg\ g^{2n}Exp\left(-\left(\alpha_k+\alpha_l\right)\left(\vec{r}_i-\vec{R}_c\right)^2g^2\right)$ are the Boys functions [S11]. It is straightforward to demonstrate that if $c_1=1$ and $c_2=c_3=c_4=c_5=0$ then based on the fact that $F_{0,ss}^i=\sqrt{\frac{\pi}{8\alpha_s}}\frac{erf\left[\sqrt{2\alpha_s}\left|\vec{r}_i-\vec{R}_c\right|\right]}{\left|\vec{r}_i-\vec{R}_c\right|}$ [S11], the effective electron-muon interaction reduces to that derived in equation (3) in the main text. Figure S2 depicts $V_{spd}^{eff}\left(\vec{r}_i\right)$ demonstrating that in contrast to $V_{e-s}^{eff}$, electrons experience a non-Coulombic anisotropic potential.

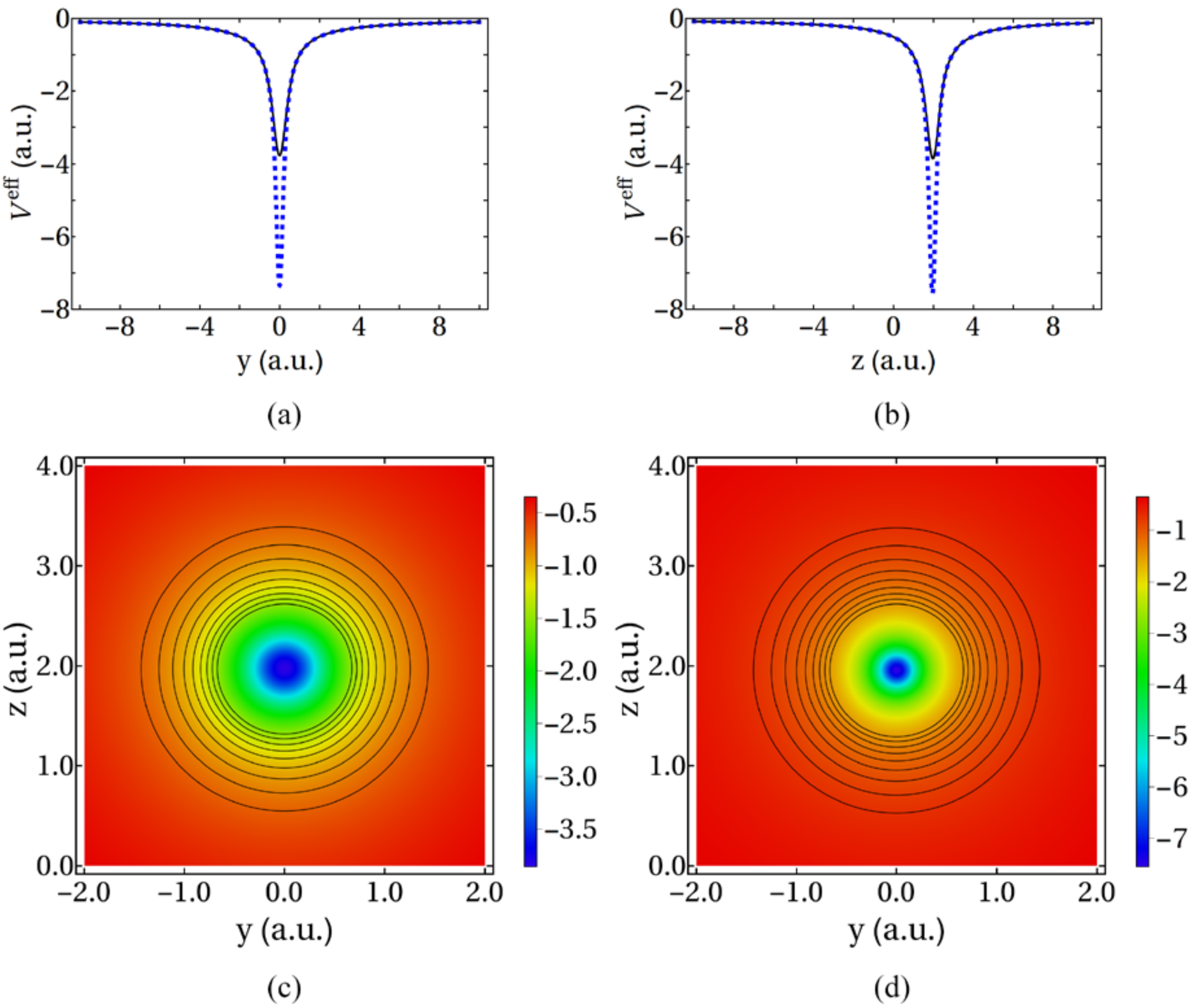

Figure S2- a) The effective muon-electron (full line) and proton-electron (dashed line) interaction potentials depicted along a y-axis, which goes through muon and is perpendicular to the z-axis. b) The same effective potentials along the z-axis. The contour maps of effective interaction potentials in μCN (c) and HCN (d) depicted at yz-plane (the contours lines are from $V_{spd}^{eff} = -0.7$ to $-1.5$, decreased in -0.1 steps). The clamped carbon nucleus is placed at the center of coordinate system while the clamped nitrogen nucleus and the banquet atom are placed at the negative and the positive sides of the z-axis, respectively.

After some mathematical manipulations, the part of the effective potential, which appears because of the kinetic energy of the muon and the muon-clamped nuclei interaction, is derived:

$$U_{spd}^{eff} = \frac{\hbar^2}{m_\mu}\{ c_{11}\left(\frac{3\alpha_s}{2}\right) + (c_{13}+c_{14}+c_{15})\left(8\sqrt{\frac{2}{3}}\right)\left(\frac{(\alpha_s\alpha_d)^{\frac{7}{4}}(3\alpha_d-2\alpha_s)}{(\alpha_s+\alpha_d)^{\frac{7}{2}}}\right) + c_{22}\left(\frac{5\alpha_p}{2}\right)$$

$$+(c_{33}+c_{44}+c_{55})\left(\frac{13\alpha_d}{6}\right) - (c_{34}+c_{35}+c_{45})\left(\frac{\alpha_d}{3}\right)\}$$

$$+\sum_{\beta}^{q} Z_{\beta}[c_{11}N_{ss}\left(\frac{2\pi}{\alpha_{ss}}\right)F_{0,ss}^{\beta}+c_{12}N_{sp}\left(\frac{4\pi}{\alpha_{sp}}\right)\overline{z}_{\beta c}F_{1,sp}^{\beta}+c_{22}N_{pp}\left(\frac{\pi}{\alpha_{pp}^{2}}\right)(F_{0,pp}^{\beta}-F_{1,pp}^{\beta}+2\alpha_{pp}\overline{z}_{\beta c}^{2}F_{2,pp}^{\beta})$$

$$+N_{sd}\left(\frac{2\pi}{\alpha_{sd}{}^{2}}\right)\{(c_{13}+c_{14}+c_{15})(F_{0,\mathrm{sd}}^{\beta}-F_{1,\mathrm{sd}}^{\beta})+2\alpha_{sd}(c_{13}\overline{x}_{\beta c}^{2}+c_{14}\overline{y}_{\beta c}^{2}+c_{15}\overline{z}_{\beta c}^{2})F_{2,\mathrm{sd}}^{\beta}\}$$

$$+N_{pd}\left(\frac{2\pi}{\alpha_{pd}^{2}}\right)\overline{z}_{\beta c}\{(c_{23}+c_{24}+3c_{25})(F_{1,pd}^{\beta}-F_{2,pd}^{\beta})+2\alpha_{pd}(c_{23}\overline{x}_{\beta c}^{2}+c_{24}\overline{y}_{\beta c}^{2}+c_{25}\overline{z}_{\beta c}^{2})F_{3,pd}^{\beta}\}$$

$$+N_{dd}\left(\frac{\pi}{2\alpha_{dd}^{3}}\right)\{3(c_{33}+c_{44}+c_{55})(F_{0,\mathrm{dd}}^{\beta}-2F_{1,\mathrm{dd}}^{\beta}+F_{2,\mathrm{dd}}^{\beta})+12\alpha_{dd}(c_{33}\overline{x}_{\beta c}^{2}+c_{44}\overline{y}_{\beta c}^{2}+c_{55}\overline{z}_{\beta c}^{2})$$
$$(F_{2,\mathrm{dd}}^{\beta}-F_{3,\mathrm{dd}}^{\beta})+4\alpha_{dd}^{2}(c_{3}\overline{x}_{\beta c}^{2}+c_{4}\overline{y}_{\beta c}^{2}+c_{5}\overline{z}_{\beta c}^{2})^{2}F_{4,\mathrm{dd}}^{\beta}+2(c_{34}+c_{35}+c_{45})(F_{0,\mathrm{dd}}^{\beta}-2F_{1,\mathrm{dd}}^{\beta}+F_{2,\mathrm{dd}}^{\beta})$$
$$+4\alpha_{dd}((c_{34}+c_{35})\overline{x}_{\beta c}^{2}+(c_{34}+c_{45})\overline{y}_{\beta c}^{2}+(c_{35}+c_{45})\overline{z}_{\beta c}^{2})(F_{2,\mathrm{dd}}^{\beta}-F_{3,\mathrm{dd}}^{\beta})\}] \quad \text{(S3)}$$

In this expression $\overline{x}_{\beta c}^{n}=\left(X_{\beta}-X_{c}\right)^{n}$, $n=0-4$ (similarly for $\overline{y}_{\beta c}^{n}$ and $\overline{z}_{\beta c}^{n}$) and $F_{n,kl}^{\beta}=\int_{0}^{1}dg\ g^{2n}Exp\left(-\left(\alpha_{k}+\alpha_{l}\right)\left(\vec{R}_{\beta}-\vec{R}_{c}\right)^{2}g^{2}\right)$, while it is straightforward to demonstrate that if $c_1=1$ and $c_2=c_3=c_4=c_5=0$ then $U_{spd}^{eff}$ reduces to $U_{s}^{eff}$. Clearly, this effective potential, $V_{spd}^{eff}=V_{e-spd}^{eff}+U_{spd}^{eff}$, is much more complicated and more reliable than the effective potential in equation (3) in the main text, yielding a new set of the EHF equations:

$$\hat{f}_{spd}^{eff}(\vec{r}_1)\psi_i(\vec{r}_1)=\varepsilon_i\psi_i(\vec{r}_1) \qquad i=1,\dots,N_e/2$$

$$\hat{f}_{spd}^{eff}(\vec{r}_1)=\hat{h}(\vec{r}_1)+V_{spd}^{eff}(\vec{r}_1)+\sum_{j}^{N_e/2}\left[2\hat{J}_j(\vec{r}_1)-\hat{K}_j(\vec{r}_1)\right]$$

$$E_{total}=E_{EHF-spd}+U_{spd}^{eff}+\sum_{\beta}^{q}\sum_{\gamma\rangle\beta}^{q}\frac{Z_{\beta}Z_{\gamma}}{\left|\vec{R}_{\beta}-\vec{R}_{\gamma}\right|} \qquad \text{(S4)}$$

The solution of the algebraic (Roothan-Hall-Hartree-Fock) version of equations (S4) using the basis set given in equation (S1) and simultaneous optimization of the muonic parameters namely, $\{c_i,\ i=1-5\}$ and $\{\alpha_k,\ k=s,p,d\}$, as well as the geometry of the clamped nuclei, $\{\vec{R}_{\beta}\}$, is completely equivalent to the solution of the NEO-HF equations and simultaneous full optimization of the parameters of the muonic [1s1p1d] basis set and the geometry of the clamped nuclei. In the case of $\mu CN$, the results given in Tables S1 and S2 are recovered from equations (S4) apart from minor differences emerging from varied numerical accuracy of the corresponding computational procedures. In the case of the muonic parameters, it is evident from Table S1 that: $c_1 > c_{t\neq 1}$, and if one starts from an initial guess where $c_1 = 1$ and $c_2 = c_3 = c_4 = c_5 = 0$, the starting effective electron-muon interaction, which is equal to that in equation (3) in the main text, varies marginally during the optimization procedure. As is also evident from Figure S3, the first term in the equation (S2), $N_{ss}\left(\frac{2\pi}{\alpha_{ss}}\right)F^{i}_{0,ss}$, is one order of magnitude larger than all the remaining terms in the electron-muon interaction and the other terms act more like perturbations modifying this dominant term.

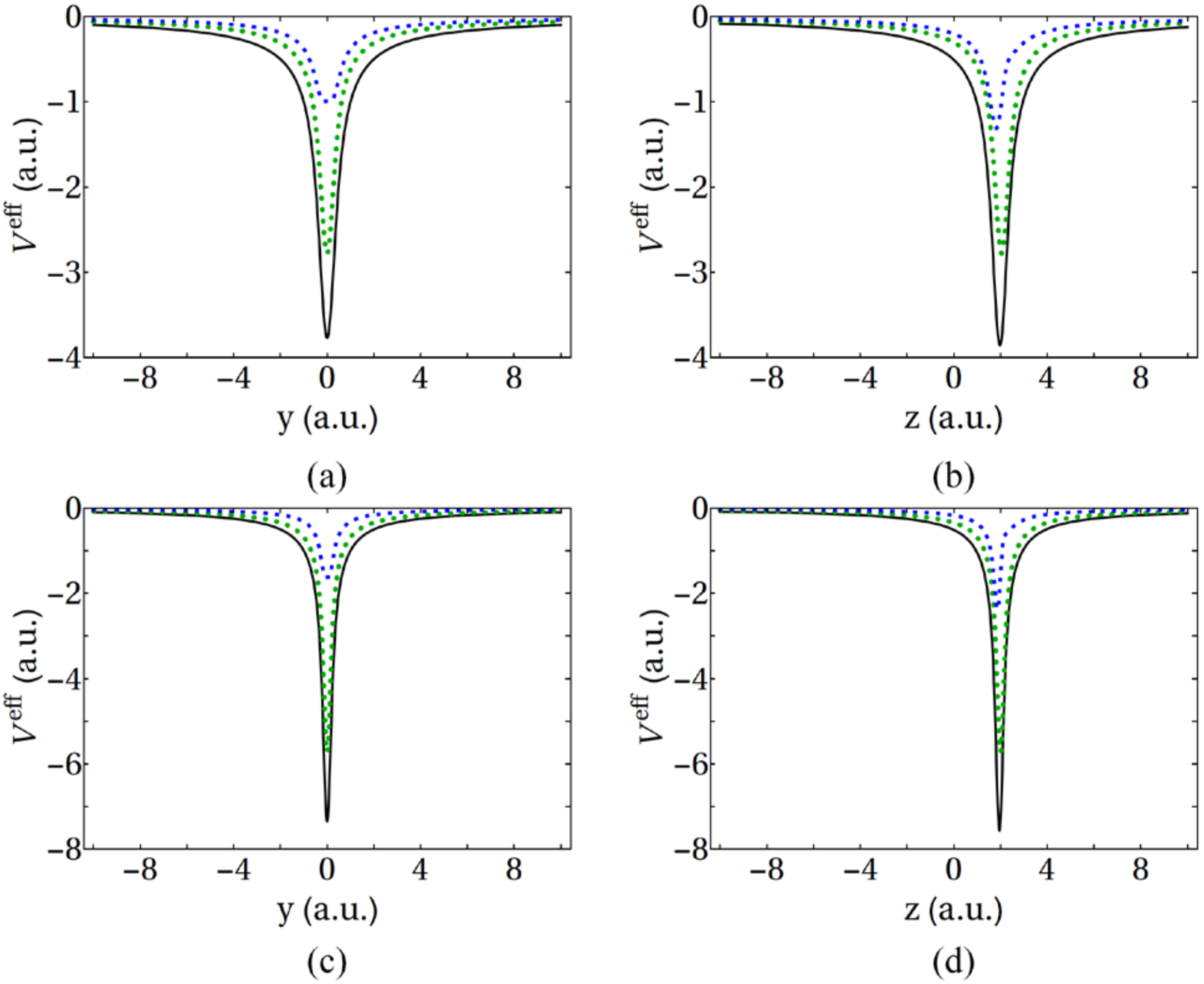

Figure S3- a) The components (the first term in equation (S2), $F_{0,ss}$, shown as green dotted, and all remaining terms, shown as blue dashed lines) and the total amount (full line) of the effective $\mu^{+}$-electron interaction potential in along a y-axis, which goes through muon and is perpendicular to the z-axis, and b) along z-axis. The same components and total amount of the effective proton-electron interaction potential along the y-axis, and (c) along the z-axis (d). The clamped carbon nucleus is placed at the center of coordinate system while the clamped nitrogen nucleus and the banquet atom are placed at the negative and the positive sides of the z-axis, respectively.

Further simplifications of the electron-muon potential are feasible based on this observation and one may derive a more compact electron-muon (or analogously muon-clamped nucleus) interaction potential simpler than equation (S2) (or equation (S3)) without a serious loss in accuracy as will be discussed in a future study.

Table S3- The optimized [1s] nuclear exponents.

| | X | Exponent | Molecule | X | Exponent |
|---|---|---|---|---|---|
| **LiX** | *T* | 37.4538 | **$NX_3$** | *T* | 43.5257 |
| | *D* | 29.9147 | | *D* | 34.7564 |
| | *H* | 20.1689 | | *H* | 23.4532 |
| | $\mu$ | 5.2910 | | $\mu$ | 6.2008 |
| **$BeX_2$** | | | **$OX_2$** | | |
| | *T* | 40.6469 | | *T* | 42.7551 |
| | *D* | 32.4777 | | *D* | 34.1756 |
| | *H* | 21.9326 | | *H* | 23.0908 |
| | $\mu$ | 5.8049 | | $\mu$ | 6.1171 |
| **$BX_3$** | | | **FX** | | |
| | *T* | 42.8437 | | *T* | 41.5458 |
| | *D* | 34.2321 | | *D* | 33.1874 |
| | *H* | 23.1255 | | *H* | 22.3832 |
| | $\mu$ | 6.1289 | | $\mu$ | 5.8789 |
| **$CX_4$** | | | | | |
| | *T* | 43.5861 | | | |
| | *D* | 34.8183 | | | |
| | *H* | 23.1255 | | | |
| | $\mu$ | 6.2158 | | | |

| | X | Exponent | Molecule | X | Exponent |
|---|---|---|---|---|---|
| **NaX** | *T* | 36.1904 | **$PX_3$** | *T* | 40.9091 |
| | *D* | 28.8930 | | *D* | 32.6743 |
| | *H* | 19.4768 | | *H* | 22.0381 |
| | $\mu$ | 5.0916 | | $\mu$ | 5.7783 |
| **$MgX_2$** | | | **$SX_2$** | | |
| | *T* | 38.2605 | | *T* | 40.2265 |
| | *D* | 30.5693 | | *D* | 32.1059 |
| | *H* | 20.6323 | | *H* | 21.6323 |
| | $\mu$ | 5.4316 | | $\mu$ | 5.6414 |
| **$AlX_3$** | | | **ClX** | | |
| | *T* | 40.0436 | | *T* | 39.0553 |
| | *D* | 31.9967 | | *D* | 31.1498 |
| | *H* | 21.6048 | | *H* | 20.9424 |
| | $\mu$ | 5.7026 | | $\mu$ | 5.4128 |
| **$SiX_4$** | | | | | |
| | *T* | 41.0820 | | | |
| | *D* | 32.8258 | | | |
| | *H* | 22.1615 | | | |
| | $\mu$ | 5.8417 | | | |

Table S4- The optimized electronic exponents of [4s1p:1s] basis set.

| | s | s | s | s | p | | s | s | s | s | p |
|---|---|---|---|---|---|---|---|---|---|---|---|
| **LiT** | 8.71 | 1.67 | 0.41 | 0.11 | 0.36 | **LiD** | 8.09 | 1.60 | 0.40 | 0.11 | 0.36 |
| **$BeT_2$** | 10.79 | 2.13 | 0.52 | 0.14 | 0.74 | **$BeD_2$** | 9.97 | 2.05 | 0.51 | 0.14 | 0.74 |
| **$BT_3$** | 12.94 | 2.65 | 0.67 | 0.18 | 1.00 | **$BD_3$** | 11.90 | 2.54 | 0.65 | 0.18 | 1.00 |
| **$CT_4$** | 14.02 | 2.94 | 0.77 | 0.22 | 1.15 | **$CD_4$** | 12.74 | 2.78 | 0.75 | 0.22 | 1.14 |
| **$NT_3$** | 14.47 | 3.06 | 0.82 | 0.26 | 0.79 | **$ND_3$** | 13.01 | 2.86 | 0.78 | 0.26 | 0.79 |
| **$OT_2$** | 9.87 | 1.86 | 0.51 | 0.15 | 0.70 | **$OD_2$** | 9.12 | 1.77 | 0.49 | 0.15 | 0.69 |
| **FT** | 10.34 | 2.00 | 0.58 | 0.16 | 0.84 | **FD** | 9.54 | 1.90 | 0.56 | 0.16 | 0.83 |
| **NaT** | 7.79 | 1.47 | 0.35 | 0.10 | 0.26 | **NaD** | 7.21 | 1.40 | 0.34 | 0.09 | 0.26 |
| **$MgT_2$** | 8.58 | 1.63 | 0.39 | 0.11 | 0.48 | **$MgD_2$** | 7.98 | 1.56 | 0.38 | 0.10 | 0.48 |
| **$AlT_3$** | 9.54 | 1.83 | 0.44 | 0.13 | 0.49 | **$AlD_3$** | 8.85 | 1.76 | 0.43 | 0.12 | 0.49 |
| **$SiT_4$** | 10.40 | 2.02 | 0.49 | 0.15 | 0.57 | **$SiD_4$** | 9.62 | 1.94 | 0.48 | 0.15 | 0.57 |
| **$PT_3$** | 10.80 | 2.13 | 0.53 | 0.17 | 0.56 | **$PD_3$** | 9.96 | 2.04 | 0.52 | 0.17 | 0.56 |
| **$ST_2$** | 11.24 | 2.24 | 0.56 | 0.18 | 0.58 | **$SD_2$** | 10.35 | 2.15 | 0.55 | 0.18 | 0.57 |
| **ClT** | 12.65 | 2.62 | 0.66 | 0.22 | 0.56 | **ClD** | 11.67 | 2.52 | 0.65 | 0.22 | 0.56 |

| | S | S | S | S | P | | S | S | S | S | P |
|---|---|---|---|---|---|---|---|---|---|---|---|
| **LiH** | 6.86 | 1.45 | 0.37 | 0.11 | 0.34 | **Liμ** | 3.54 | 0.97 | 0.29 | 0.09 | 0.31 |
| **$BeH_2$** | 8.53 | 1.88 | 0.49 | 0.13 | 0.74 | **Beμ$_2$** | 4.41 | 1.31 | 0.41 | 0.12 | 0.73 |
| **$BH_3$** | 10.16 | 2.35 | 0.63 | 0.18 | 0.98 | **Bμ$_3$** | 5.16 | 1.65 | 0.53 | 0.16 | 0.88 |
| **$CH_4$** | 10.55 | 2.49 | 0.71 | 0.21 | 1.10 | **Cμ$_4$** | 3.60 | 0.89 | 0.27 | 0.07 | 0.95 |
| **$NH_3$** | 10.49 | 2.46 | 0.72 | 0.24 | 0.77 | **Nμ$_3$** | 4.02 | 1.05 | 0.34 | 0.10 | 0.69 |
| **$OH_2$** | 7.89 | 1.63 | 0.47 | 0.14 | 0.69 | **Oμ$_2$** | 4.45 | 1.21 | 0.40 | 0.12 | 0.65 |
| **FH** | 8.16 | 1.73 | 0.53 | 0.15 | 0.83 | **Fμ** | 3.86 | 1.01 | 0.35 | 0.10 | 0.79 |
| **NaH** | 6.19 | 1.28 | 0.32 | 0.09 | 0.25 | **Naμ** | 3.24 | 0.86 | 0.25 | 0.08 | 0.23 |
| **$MgH_2$** | 6.90 | 1.45 | 0.36 | 0.10 | 0.48 | **Mgμ$_2$** | 3.73 | 1.04 | 0.30 | 0.09 | 0.46 |
| **$AlH_3$** | 7.63 | 1.62 | 0.41 | 0.12 | 0.48 | **Alμ$_3$** | 4.07 | 1.15 | 0.34 | 0.11 | 0.43 |
| **$SiH_4$** | 8.28 | 1.79 | 0.46 | 0.14 | 0.56 | **Siμ$_4$** | 4.43 | 1.28 | 0.39 | 0.13 | 0.52 |
| **$PH_3$** | 8.49 | 1.86 | 0.49 | 0.16 | 0.55 | **Pμ$_3$** | 4.27 | 1.24 | 0.39 | 0.15 | 0.47 |
| **$SH_2$** | 8.85 | 1.98 | 0.53 | 0.18 | 0.56 | **Sμ$_2$** | 4.55 | 1.36 | 0.44 | 0.16 | 0.49 |
| **ClH** | 9.88 | 2.33 | 0.62 | 0.22 | 0.55 | **Clμ** | 5.61 | 1.82 | 0.55 | 0.20 | 0.49 |

Table S5- The optimized muonic exponents of [4s1p:2s2p2d] basis set.

| | s | s | p | p | d | d |
|---|---|---|---|---|---|---|
| **Liμ** | 7.52 | 7.00 | 5.53 | 4.01 | 6.59 | 4.63 |
| **$Be\mu_2$** | 8.28 | 6.34 | 4.43 | 3.79 | 6.57 | 4.75 |
| **$B\mu_3$** | 8.00 | 6.83 | 5.60 | 3.82 | 6.84 | 4.77 |
| **$C\mu_4$** | 8.93 | 7.66 | 6.01 | 4.11 | 7.18 | 4.98 |
| **$N\mu_3$** | 8.14 | 7.34 | 6.54 | 4.40 | 7.31 | 4.77 |
| **$O\mu_2$** | 9.58 | 5.81 | 6.94 | 4.49 | 7.68 | 4.64 |
| **Fμ** | 11.37 | 5.38 | 7.16 | 4.30 | 5.45 | 4.05 |
| | | | | | | |
| **Naμ** | 7.56 | 7.08 | 5.63 | 3.95 | 6.22 | 4.42 |
| **$Mg\mu_2$** | 7.72 | 7.00 | 5.97 | 4.41 | 6.50 | 4.60 |
| **$Al\mu_3$** | 7.54 | 6.63 | 5.80 | 4.45 | 6.40 | 4.53 |
| **$Si\mu_4$** | 7.87 | 6.75 | 5.66 | 4.23 | 6.56 | 4.64 |
| **$P\mu_3$** | 8.51 | 6.75 | 6.13 | 4.34 | 6.63 | 4.67 |
| **$S\mu_2$** | 7.44 | 6.75 | 6.35 | 4.29 | 6.66 | 4.46 |
| **Clμ** | 7.38 | 6.63 | 6.22 | 4.06 | 6.63 | 4.29 |

Table S6- The optimized electronic exponents of [4s1p:2s2p2d] basis set.

| | s | s | s | s | p | | s | s | s | s | p |
|---|---|---|---|---|---|---|---|---|---|---|---|
| **LiT** | 8.74 | 1.67 | 0.41 | 0.11 | 0.29 | **LiD** | 8.06 | 1.60 | 0.40 | 0.11 | 0.29 |
| **$BeT_2$** | 10.80 | 2.14 | 0.52 | 0.14 | 0.68 | **$BeD_2$** | 10.47 | 2.21 | 0.57 | 0.16 | 0.12 |
| **$BT_3$** | 12.91 | 2.65 | 0.67 | 0.18 | 0.91 | **$BD_3$** | 11.89 | 2.54 | 0.65 | 0.18 | 0.90 |
| **$CT_4$** | 13.57 | 2.85 | 0.75 | 0.22 | 1.00 | **$CD_4$** | 12.19 | 2.66 | 0.72 | 0.22 | 0.98 |
| **$NT_3$** | 14.30 | 3.12 | 0.85 | 0.27 | 0.61 | **$ND_3$** | 12.91 | 2.93 | 0.82 | 0.26 | 0.60 |
| **$OT_2$** | 9.16 | 1.75 | 0.48 | 0.14 | 0.59 | **$OD_2$** | 8.59 | 1.70 | 0.47 | 0.14 | 0.59 |
| **FT** | 9.61 | 1.92 | 0.56 | 0.15 | 0.71 | **FD** | 8.94 | 1.85 | 0.55 | 0.15 | 0.71 |
| | | | | | | | | | | | |
| **NaT** | 7.84 | 1.47 | 0.35 | 0.10 | 0.23 | **NaD** | 7.25 | 1.41 | 0.34 | 0.09 | 0.22 |
| **$MgT_2$** | 8.61 | 1.63 | 0.39 | 0.11 | 0.45 | **$MgD_2$** | 8.57 | 1.73 | 0.43 | 0.11 | 0.14 |
| **$AlT_3$** | 10.18 | 2.03 | 0.51 | 0.15 | 0.08 | **$AlD_3$** | 9.40 | 1.94 | 0.49 | 0.15 | 0.08 |
| **$SiT_4$** | 10.37 | 2.02 | 0.49 | 0.15 | 0.51 | **$SiD_4$** | 9.61 | 1.94 | 0.48 | 0.15 | 0.51 |
| **$PT_3$** | 10.72 | 2.13 | 0.53 | 0.17 | 0.49 | **$PD_3$** | 9.90 | 2.04 | 0.52 | 0.17 | 0.48 |
| **$ST_2$** | 11.05 | 2.24 | 0.57 | 0.18 | 0.49 | **$SD_2$** | 10.20 | 2.15 | 0.56 | 0.18 | 0.49 |
| **ClT** | 12.18 | 2.62 | 0.67 | 0.22 | 0.47 | **ClD** | 11.24 | 2.52 | 0.66 | 0.22 | 0.47 |

| | s | s | s | s | p | | s | s | s | s | p |
|---|---|---|---|---|---|---|---|---|---|---|---|
| **LiH** | 6.90 | 1.46 | 0.37 | 0.11 | 0.28 | **Liμ** | 3.54 | 0.96 | 0.29 | 0.09 | 0.23 |
| **$BeH_2$** | 8.84 | 2.01 | 0.54 | 0.15 | 0.12 | **$Be\mu_2$** | 4.43 | 1.31 | 0.41 | 0.12 | 0.63 |
| **$BH_3$** | 10.13 | 2.34 | 0.63 | 0.18 | 0.87 | **$B\mu_3$** | 5.07 | 1.60 | 0.52 | 0.16 | 0.73 |
| **$CH_4$** | 10.30 | 2.41 | 0.69 | 0.21 | 0.94 | **$C\mu_4$** | 3.63 | 0.91 | 0.28 | 0.07 | 0.77 |
| **$NH_3$** | 10.35 | 2.50 | 0.75 | 0.25 | 0.59 | **$N\mu_3$** | 4.07 | 1.08 | 0.34 | 0.10 | 0.54 |
| **$OH_2$** | 7.50 | 1.58 | 0.45 | 0.13 | 0.58 | **$O\mu_2$** | 4.40 | 1.23 | 0.40 | 0.12 | 0.54 |
| **FH** | 7.79 | 1.73 | 0.53 | 0.15 | 0.70 | **Fμ** | 4.24 | 1.33 | 0.48 | 0.13 | 0.64 |
| | | | | | | | | | | | |
| **NaH** | 6.23 | 1.29 | 0.32 | 0.09 | 0.22 | **Naμ** | 3.28 | 0.87 | 0.25 | 0.08 | 0.19 |
| **$MgH_2$** | 7.38 | 1.59 | 0.41 | 0.11 | 0.14 | **$Mg\mu_2$** | 3.77 | 1.04 | 0.30 | 0.09 | 0.41 |
| **$AlH_3$** | 8.05 | 1.78 | 0.47 | 0.14 | 0.08 | **$Al\mu_3$** | 4.11 | 1.16 | 0.34 | 0.11 | 0.36 |
| **$SiH_4$** | 8.27 | 1.80 | 0.46 | 0.14 | 0.49 | **$Si\mu_4$** | 4.43 | 1.28 | 0.38 | 0.13 | 0.43 |
| **$PH_3$** | 8.47 | 1.87 | 0.50 | 0.16 | 0.46 | **$P\mu_3$** | 4.30 | 1.26 | 0.40 | 0.15 | 0.38 |
| **$SH_2$** | 8.74 | 1.99 | 0.54 | 0.18 | 0.47 | **$S\mu_2$** | 4.30 | 1.31 | 0.44 | 0.16 | 0.38 |
| **ClH** | 6.31 | 1.96 | 0.58 | 0.21 | 0.49 | **Clμ** | 5.47 | 1.88 | 0.59 | 0.20 | 0.39 |

Table S7- Total energies computed with the optimized and averaged exponents using [6-311+g(d)/4s1p:1s] and [6-311+g(d)/4s1p:2s2p2d] basis sets. The energy differences between the optimized and averaged basis sets have been given in columns with the headline "Diff." in milli-Hartrees. The two last columns contain the energy difference between the two averaged basis sets and between the two optimized basis sets in milli-Hartrees.

| | **1s Opt.** | **1s Ave.** | **1s Diff.** | **2s2p2d Opt.** | **2s2p2d Ave.** | **2s2p2d Diff.** | **Ave. Diff.** | **Opt. Diff.** |
|---|---|---|---|---|---|---|---|---|
| **Liμ** | -7.89187 | -7.89149 | 0.38 | -7.89199 | -7.89178 | 0.21 | 0.29 | 0.12 |
| **Beμ2** | -15.56610 | -15.56603 | 0.07 | -15.56657 | -15.56651 | 0.05 | 0.49 | 0.47 |
| **Bμ3** | -26.07424 | -26.07322 | 1.02 | -26.07575 | -26.07532 | 0.43 | 2.09 | 1.51 |
| **Cμ4** | -39.77161 | -39.76978 | 1.83 | -39.77534 | -39.77454 | 0.80 | 4.76 | 3.73 |
| **Nμ3** | -55.88687 | -55.88607 | 0.80 | -55.89250 | -55.89232 | 0.17 | 6.25 | 5.63 |
| **Oμ2** | -75.83795 | -75.83754 | 0.41 | -75.84431 | -75.84408 | 0.24 | 6.53 | 6.36 |
| **Fμ** | -99.94933 | -99.94847 | 0.86 | -99.95406 | -99.95334 | 0.72 | 4.87 | 4.73 |
| | | | | | | | | |
| **Naμ** | -162.28828 | -162.28736 | 0.92 | -162.28838 | -162.28783 | 0.55 | 0.47 | 0.10 |
| **Mgμ2** | -200.53838 | -200.53790 | 0.48 | -200.53866 | -200.53839 | 0.27 | 0.49 | 0.27 |
| **Alμ3** | -243.33596 | -243.33570 | 0.25 | -243.33676 | -243.33656 | 0.20 | 0.86 | 0.80 |
| **Siμ4** | -290.84045 | -290.84030 | 0.15 | -290.84212 | -290.84204 | 0.09 | 1.74 | 1.68 |
| **Pμ3** | -342.17069 | -342.17018 | 0.50 | -342.17293 | -342.17243 | 0.50 | 2.24 | 2.24 |
| **Sμ2** | -398.50143 | -398.50105 | 0.38 | -398.50420 | -398.50380 | 0.40 | 2.75 | 2.78 |
| **Clμ** | -459.99891 | -459.99857 | 0.34 | -460.00098 | -460.00073 | 0.26 | 2.16 | 2.08 |

Table S8- The distances between the banquet atoms and the central clamped nuclei computed with the optimized and averaged exponents using [6-311+g(d)/4s1p:1s] and [6-311+g(d)/4s1p:2s2p2d] basis sets. The distance differences between the optimized and averaged basis sets have been given in columns with the headline “Diff.” in Angstroms. The two last columns contain the distance difference between the two averaged basis sets and between the two optimized basis sets in Angstroms.

| | **1s Opt.** | **1s Ave.** | **1s Diff.** | **2s2p2d Opt.** | **2s2p2d Ave.** | **2s2p2d Diff.** | **Ave. Diff.** | **Opt. Diff.** |
|---|---|---|---|---|---|---|---|---|
| **Liμ** | 1.697 | 1.688 | 0.009 | 1.683 | 1.686 | -0.003 | 0.002 | 0.014 |
| **Beμ2** | 1.415 | 1.416 | -0.001 | 1.401 | 1.395 | 0.005 | 0.021 | 0.015 |
| **Bμ3** | 1.267 | 1.273 | -0.006 | 1.240 | 1.234 | 0.006 | 0.039 | 0.027 |
| **Cμ4** | 1.155 | 1.161 | -0.006 | 1.119 | 1.115 | 0.005 | 0.047 | 0.036 |
| **Nμ3** | 1.068 | 1.073 | -0.005 | 1.017 | 1.016 | 0.001 | 0.057 | 0.051 |
| **Oμ2** | 1.006 | 1.010 | -0.003 | 0.938 | 0.937 | 0.001 | 0.073 | 0.069 |
| **Fμ** | 0.964 | 0.966 | -0.002 | 0.842 | 0.848 | -0.006 | 0.118 | 0.122 |
| | | | | | | | | |
| **Naμ** | 2.000 | 1.986 | 0.013 | 1.987 | 1.990 | -0.002 | -0.003 | 0.012 |
| **Mgμ2** | 1.793 | 1.789 | 0.005 | 1.782 | 1.784 | -0.002 | 0.005 | 0.012 |
| **Alμ3** | 1.666 | 1.664 | 0.002 | 1.647 | 1.648 | -0.001 | 0.016 | 0.020 |
| **Siμ4** | 1.561 | 1.562 | 0.000 | 1.536 | 1.536 | 0.000 | 0.026 | 0.026 |
| **Pμ3** | 1.492 | 1.491 | 0.001 | 1.453 | 1.456 | -0.003 | 0.035 | 0.038 |
| **Sμ2** | 1.411 | 1.408 | 0.003 | 1.353 | 1.359 | -0.006 | 0.049 | 0.058 |
| **Clμ** | 1.349 | 1.344 | 0.005 | 1.276 | 1.283 | -0.007 | 0.061 | 0.073 |

Table S9- The angles between the two banquet atoms through the central clamped nuclei computed with the optimized and averaged exponents using [6-311+g(d)/4s1p:1s] and [6-311+g(d)/4s1p:2s2p2d] basis sets. The angle differences between the optimized and averaged basis sets have been given in columns with the headline "Diff." in degrees. The two last columns contain the angle difference between the two averaged basis sets and between the two optimized basis sets in degrees.

| | **1s Opt.** | **1s Ave.** | **1s Diff.** | **2s2p2d Opt.** | **2s2p2d Ave.** | **2s2p2d Diff.** | **Ave. Diff.** | **Opt. Diff.** |
|---|---|---|---|---|---|---|---|---|
| **$N\mu_3$** | 109.1 | 108.5 | 0.6 | 109.1 | 108.8 | 0.3 | -0.3 | 0.0 |
| **$O\mu_2$** | 107.6 | 107.3 | 0.3 | 107.4 | 107.4 | -0.1 | -0.1 | 0.2 |
| **$P\mu_3$** | 95.1 | 95.3 | -0.2 | 95.0 | 95.1 | 0.0 | 0.2 | 0.1 |
| **$S\mu_2$** | 93.9 | 94.1 | -0.2 | 93.9 | 93.7 | 0.1 | 0.4 | 0.1 |

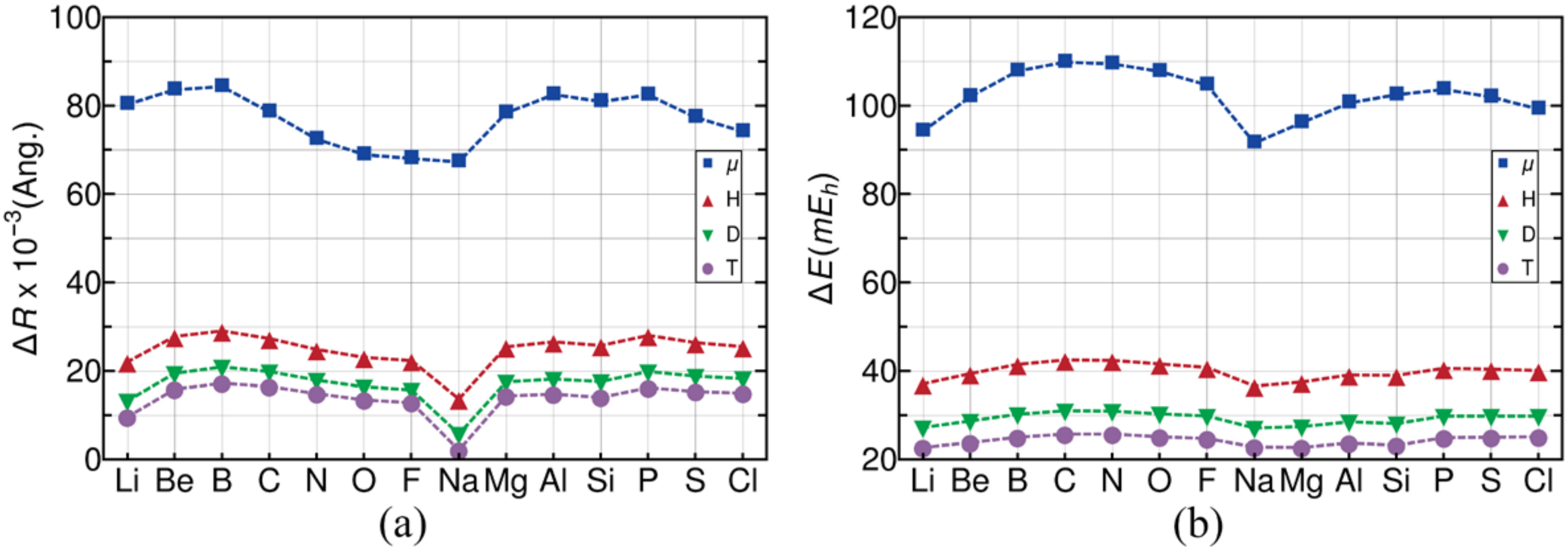

**Figure S4-** The difference in the mean inter-nuclear distances (of the quantum nucleus and the central atom distance) (a) and the difference in total energies (b) of the singly-substituted X= $\mu$, H, D, T species relative to their clamped nucleus counterparts, computed at NEO-HF/[6-311++g(d,p)/4s1p:1s] and HF/6-311++g(d,p) levels, respectively.